\documentclass[12pt]{iopart}

\usepackage{graphicx}
\usepackage{cite}
\usepackage{epsfig}
\usepackage{svg}
\makeatletter
\expandafter\let\csname equation*\endcsname\relax
\expandafter\let\csname endequation*\endcsname\relax
\makeatother
\usepackage{amsmath}
\usepackage{comment}
\usepackage{adjustbox}
\usepackage[center]{caption}\usepackage{hyperref}
\usepackage{listings}
\usepackage{booktabs}
\usepackage{siunitx}
\usepackage{subcaption}
\usepackage{multirow}
\usepackage{float}

\usepackage{pifont}
\usepackage{colortbl}

\begin{document}

\title[Hardware-Aware SNN Design Framework]{A Hardware-Aware Open-Source Framework for Design Space Exploration of Mixed-Signal Spiking Neural Networks}
\author{
Sayma Nowshin Chowdhury$^{1,\dagger}$,
Vineeta Nair$^{1,\dagger}$,
Taseen Forhad$^{1,\dagger}$,
Aishwarya Natarajan$^{2}$,
Corey Hart$^{3}$,
Sahil Shah$^{1,*}$
}

\address{$^1$ Department of Electrical and Computer Engineering, University of Maryland, College Park, MD, USA}
\address{$^2$ Hewlett Packard Labs, USA}
\address{$^3$ Lockheed Martin, USA}

\ead{[sshah389@umd.edu](mailto:sshah389@umd.edu)}

\begin{indented}
\item[]$^{\dagger}$ These authors contributed equally to this work.
\item[]$^{*}$ Corresponding author.
\item[]June 2026
\end{indented}


\begin{abstract}
Energy-efficient neuromorphic computing at the edge requires simulation tools that can capture the non-ideal behavior of mixed-signal spiking neural network (SNN) hardware while supporting system-level design exploration. This work presents an open-source hardware-aware simulation framework for mixed-signal SNNs that enables comparative analysis across neuron, synapse and architecture choices. The framework supports multiple neuron models, including Leaky Integrate-and-Fire (LIF), Hodgkin-Huxley (HH) and Axon-Hillock (AH), together with non-volatile analog synapses based on floating-gate transistors and ReRAM devices. By incorporating device-level nonlinearities directly into PyTorch-based training and inference, the tool enables optimization of physical synaptic parameters rather than idealized abstract weights. The framework is evaluated on standard neuromorphic benchmarks, including N-MNIST, DVS Gesture and Spiking Heidelberg Digits (SHD). For each model dataset configuration, it reports classification accuracy together with hardware-oriented metrics such as silicon area, power consumption and quantization sensitivity. These capabilities enable cross-layer design space exploration and help identify neuron-synapse configurations that best satisfy application-specific constraints on accuracy, energy efficiency, area and hardware fidelity.
\end{abstract}

\section{Introduction}
Edge devices support a wide range of applications, including remote monitoring and sensing, continuous vital-sign and physiological-signal monitoring and autonomous ground and aerial robots operating in unknown environments~\cite{baek_edge_2025}. The data generated by these sensors are increasingly processed using machine learning algorithms. However, edge devices are severely constrained by size, weight, area and power (SWAP), motivating the exploration of new circuit and system architectures that can efficiently process sensor data at the edge.

Neuromorphic architectures, particularly spiking neural networks (SNNs), offer a promising approach for energy-efficient edge intelligence by drawing inspiration from the event-driven operation of the brain. Unlike conventional rate-encoded neural networks, SNNs operate through discrete spike events that closely mirror the temporal dynamics of biological neurons, offering inherent advantages in energy efficiency and temporal information processing \cite{Maass1997,schuman_survey_2017}. This event-driven computation model becomes particularly compelling when implemented on neuromorphic hardware platforms, where the spatially co-located memory and compute elements enable significantly higher energy efficiency compared to traditional von Neumann architectures\cite{davies_loihi_2018,akopyan_truenorth_2015,painkras_spinnaker_2013,park_65k-neuron_2014}. The biological plausibility of SNNs extends beyond their spike-based communication to encompass local learning mechanisms such as spike-timing dependent plasticity (STDP) \cite{caporale2008spike}, which demonstrate substantially lower memory requirements and power consumption compared to conventional error backpropagation algorithms.

\begin{figure}[htbp]
\centering
\includegraphics[width=0.9\linewidth]{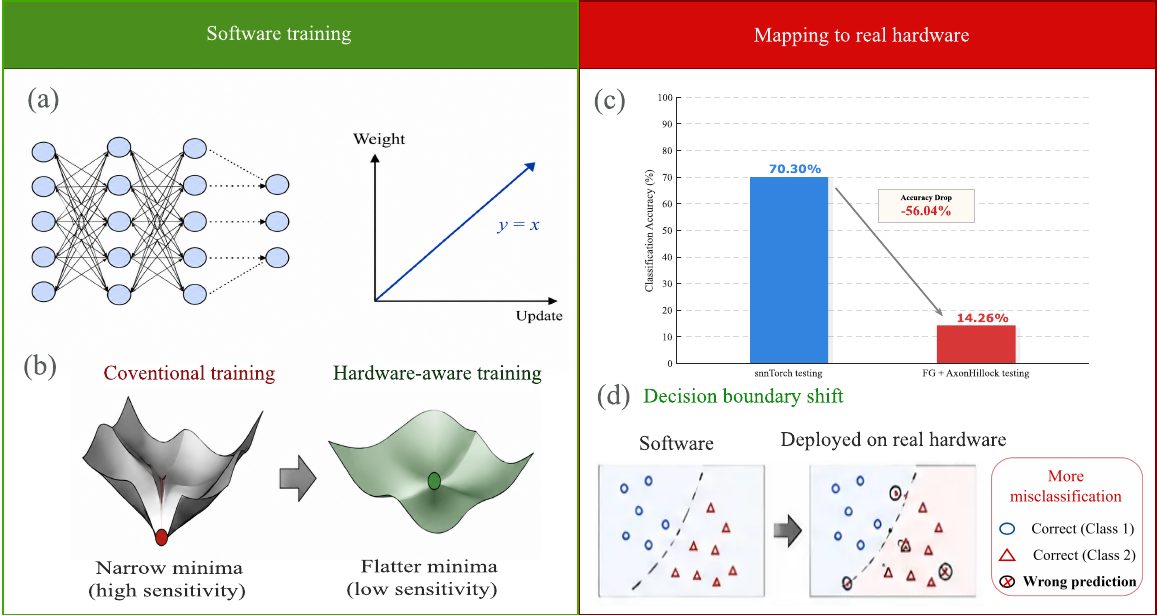}
\caption{Impact of hardware non-idealities on neuromorphic learning and inference: (a) Software training assumes ideal linear weight behavior, also (b) assumed standard (software) training leads to narrow minima sensitive to hardware mismatches and nonlinearities. Consequently, (c) direct deployment of SnnTorch-trained weights onto the floating-gate and Axon Hillock hardware model produces accuracy degradation in the SHD task due to (d) shifts in the decision boundary and a corresponding increase in misclassification rate.}
\label{fig01}
\end{figure}

eThe implementation of SNNs on mixed-signal neuromorphic hardware presents both significant opportunities and substantial challenges. Mixed-signal circuits have demonstrated superior energy efficiency compared with digital implementations \cite{mead_neuromorphic_1990,indiveri_memory_2015,qiao2015reconfigurable,chatterjee2019exploiting,rubino2023neuromorphic}; however, they are inherently affected by nonlinearities, device mismatch, and process variations that can degrade network performance \cite{11412426}. These hardware non-idealities  introduce a fundamental gap between the theoretical performance of SNN algorithms developed in software and their practical deployment on analog neuromorphic circuits. This gap is quantitatively illustrated in Fig.~\ref{fig01}(c), which demonstrates an accuracy degradation of 56.04\% incurred when weights trained using snnTorch~\cite{eshraghian_training_2023-1} are directly mapped onto a mixed-signal neuromorphic SNN without hardware-aware compensation. Hence, conventional training can converge to solutions that are more sensitive to parameter perturbations, whereas hardware-aware training encourages flatter minima that are inherently more robust to weight variations and non-idealities, consistent with the observed benefits of wide flat minima in neural network optimization (Fig.~\ref{fig01}(b))\cite{baldassi2020shaping}. As a consequence, the conventional approach of compiling software-trained parameters directly onto mixed-signal hardware often requires extensive calibration and fine-tuning of on-chip biases and voltages, increasing engineering effort and limiting the accessibility and scalability of neuromorphic computing solutions. This disconnect between algorithm development and hardware realization represents a critical bottleneck in the design of deployable neuromorphic systems.

To address these challenges, there is a critical need for design space exploration tools that can accurately capture the interaction between analog hardware characteristics and learning dynamics in mixed-signal SNNs. Existing methodologies often lack comprehensive frameworks capable of systematically evaluating the trade-offs among classification accuracy, power consumption, and silicon area while accounting for realistic circuit-level constraints and device non-idealities.

This work presents a hardware-aware simulation and design exploration framework for mixed-signal SNNs that incorporates analog synapse models based on floating-gate (FG) transistors and ReRAM (Resistive Random-Access Memory) devices, together with mixed-signal neuron implementations. By embedding device-level behavior directly into the learning and inference process, the framework enables realistic evaluation of neuromorphic architectures and helps bridge the gap between algorithm-level SNN research and practical hardware deployment.

The primary contribution of this work is the development of a unified hardware-aware simulation framework for mixed-signal SNNs that bridges device-level neuromorphic hardware modeling with network-level learning and evaluation. The framework integrates experimentally calibrated FG transistor (65 nm) and ReRAM synapse models (130 nm), together with multiple mixed-signal neuron implementations, including Axon-Hillock (AH), standard  Leaky Integrate-and-Fire (LIF), adaptive LIF, Schmitt-triggered LIF, and Hodgkin--Huxley (HH) neurons, validated against transistor-level Cadence simulations. By embedding nonlinear device characteristics directly into custom PyTorch-based training modules, the framework enables direct optimization of physical synaptic parameters such as floating-gate voltage and ReRAM filament gap, within the SNN training loop, rather than relying on idealized abstract weights, thereby eliminating the need for post-training weight mapping. In addition, the tool supports both fully connected and recurrent SNN architectures and enables evaluation across standard neuromorphic benchmark datasets, including N-MNIST, DVS Gesture, and Spiking Heidelberg Digits (SHD). Finally, the framework incorporates hardware-oriented analysis of classification accuracy, silicon area, power consumption, and synaptic quantization effects, enabling systematic quantitative design space exploration for mixed-signal neuromorphic systems.

\section{Background and Related Work}

Over the past few decades, numerous simulation frameworks have been developed to emulate spiking neural dynamics and evaluate large-scale neuromorphic systems. These tools have played a crucial role in bridging algorithmic neuroscience and computational modeling, enabling the exploration of spike-based computation, synaptic plasticity and network connectivity in reproducible environments~\cite{Gerstner2014, Maass1997}. Table~\ref{tab:prior_works} summarizes representative neuromorphic simulation platforms that span a wide range of abstraction levels, from detailed biophysical neuron models to large-scale event-driven architectures optimized for scalability and performance. Each framework adopts distinct modeling strategies to balance biological realism, computational efficiency, and simulation tractability.

\begin{table}[h!]
\centering
\caption{Comparison with Prior works on SNN modeling}
\label{tab:prior_works}

\renewcommand{\arraystretch}{1.2}

\scalebox{0.65}{
\begin{tabular}{|p{6cm}|p{7cm}|p{10cm}|}
\hline

\textbf{Work} & \textbf{Neuron model} & \textbf{Key contributions} \\

\hline
GENESIS \cite{wilson1988genesis} 
& HH  
& Modular biologically realistic neural network simulator with hierarchical organization.  \\
\hline

NEURON \cite{10.1162/neco.1997.9.6.1179} 
& HH  
& Compartmental neural modeling using cable equation formalism for detailed biophysical simulations. \\
\hline

XPPAUT \cite{Ermentrout2002} 
& IF   
& General-purpose dynamical systems solver for neuronal differential equation analysis. \\
\hline

Circuit simulator \cite{Natschlager2003} 
& HH, LIF, Izhikevich IF, compartmental neurons  
& MATLAB-based framework supporting multiple neuron and synapse models including static and dynamic synapses. \\
\hline

Synaptic conductance model \cite{brette2006exact} 
& IF 
& Event-driven simulation of integrate-and-fire neurons using polynomial root-finding (Sturm sequences). \\
\hline

NEST \cite{Gewaltig2007} 
& HH, LIF, Izhikevich IF  
& Minimal computational synapse abstraction enabling scalable, low-overhead, event-driven large-scale neural simulation. \\
\hline

Voltage stepping model \cite{kaabi2011performance} 
& Adaptive IF 
& Voltage-stepping scheme extending event-driven simulation to nonlinear adaptive neuron models. \\
\hline

SPAUN \cite{Eliasmith2012} 
& HH  
& Semantic pointer architecture enabling algebraic manipulation of compressed neural representations. \\
\hline

NeoCortical simulator \cite{Hoang2003} 
& HH, LIF  
& Conductance-based synapse modeling with scalable CPU/GPU large-scale neural simulation support. \\
\hline

MOSAIC \cite{dalgaty2024mosaic} 
& LIF 
& Energy-efficient edge inference architecture using ReRAM-based synaptic computation. \\
\hline

\textbf{This work} 
& \textbf{AH, LIF, HH} 
& \textbf{Incorporates FG and ReRAM synapse models with  non-ideal device physics for neuromorphic exploration.} \\
\hline

\end{tabular}
}
\vspace{0.4em}
\\
\begin{minipage}{\textwidth}
        \footnotesize
        HH - Hodgkin–Huxley , IF - Integrate-and-Fire, LIF - Leaky Integrate-and-Fire, FG- floating-gate
 transistors, ReRAM - Resistive Random-Access Memory
    \end{minipage}
\end{table}

From an algorithmic perspective, SNN simulators typically follow either \textit{synchronous} (clock-driven) or \textit{asynchronous} (event-driven) execution paradigms. Clock-driven approaches update all neurons and synapses at fixed time steps, offering simplicity and flexibility across neuron models at the cost of discretized spike timing~\cite{Carnevale2006}. In contrast, event-driven methods update states only upon spike events, achieving higher temporal precision but requiring complex scheduling mechanisms and often restricting the choice of neuron models to those with analytically tractable dynamics~\cite{Brette2007}.

Widely used frameworks such as \textbf{GENESIS}~\cite{wilson1988genesis} and \textbf{NEURON}~\cite{10.1162/neco.1997.9.6.1179} have enabled detailed modeling of conductance-based neurons and dendritic structures, supporting cellular-level investigations with high biological fidelity. At the other end of the spectrum, large-scale simulators such as \textbf{NEST}~\cite{Gewaltig2007} and related platforms have been optimized for scalability, enabling the simulation of networks with millions of neurons through efficient event-driven computation and distributed architectures. Other tools, including \textbf{CSIM}~\cite{Natschlager2003} and \textbf{XPPAUT}~\cite{Ermentrout2002}, provide flexible environments for rapid prototyping and dynamical systems analysis.

\begin{figure}[htbp]
\centering
\includegraphics[scale=0.9]{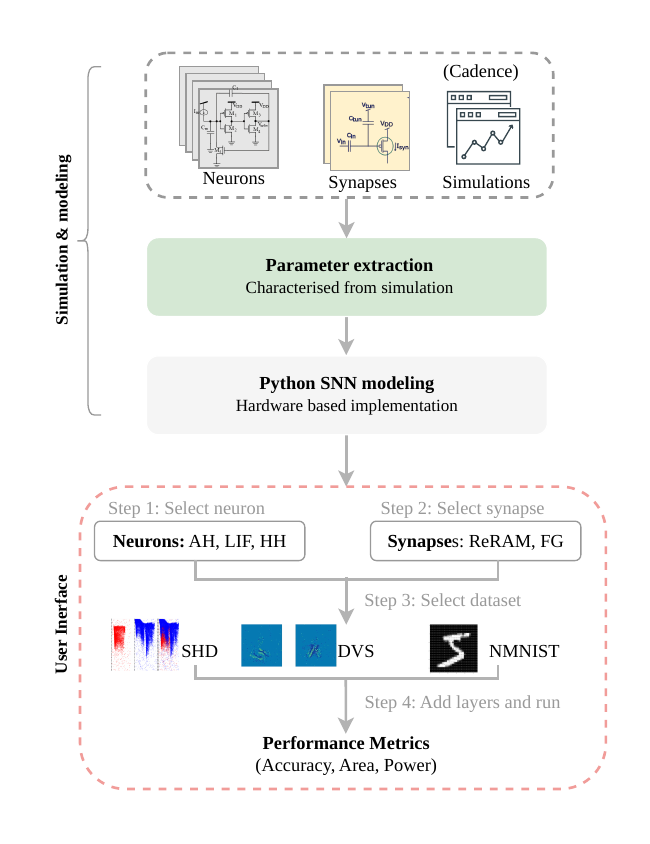}
\caption{General flow of the proposed simulation framework for mixed-signal SNNs. The tool integrates hardware-aware neuron and synapse models with learning algorithms, enabling design space exploration across accuracy, power and area metrics.}
\label{fig02}
\end{figure}

Despite their success, a fundamental limitation of existing frameworks is their abstraction away from physical hardware constraints. Most simulators assume idealized neuron and synapse behavior, neglecting the impact of transistor-level effects such as device mismatch, parasitic capacitances, leakage currents and supply-voltage scaling. Furthermore, synaptic models are often simplified as linear or instantaneous weight updates, overlooking the inherently nonlinear and state-dependent characteristics of analog and non-volatile devices such as FG and ReRAM. As a result, there exists a significant gap between algorithmic SNN development and its deployment on mixed-signal neuromorphic hardware, where these non-idealities directly influence network performance, stability and energy efficiency.

To address this gap, this work introduces a hardware-aware simulation framework that embeds technology-dependent circuit constraints directly into neuron and synapse models. As illustrated in Fig.~\ref{fig02}, the proposed tool enables cross-layer co-design by jointly modeling device physics, circuit behavior and network-level learning dynamics. Neuron models are parameterized using circuit-level quantities such as biasing conditions and membrane capacitance, ensuring compatibility with advanced CMOS technology nodes (e.g., 28 nm and 65 nm). In parallel, synaptic weights are implemented using analog non-volatile devices, including FG and ReRAM, whose nonlinear characteristics are incorporated into the learning process. 

By explicitly capturing the interaction between hardware non-idealities and learning dynamics, the proposed framework enables realistic evaluation of mixed-signal SNNs and facilitates informed design space exploration. This capability is essential for translating the theoretical advantages of SNNs into practical, energy-efficient neuromorphic hardware systems.
\section{Spiking Neural Network Primitives}

Spiking Neural Networks (SNNs) represent a class of neural architectures that process information using discrete, time-dependent events known as spikes, inspired by biological neural systems~\cite{Maass1997}. 
\begin{figure}[htbp]
\begin{center}
\epsfig{file=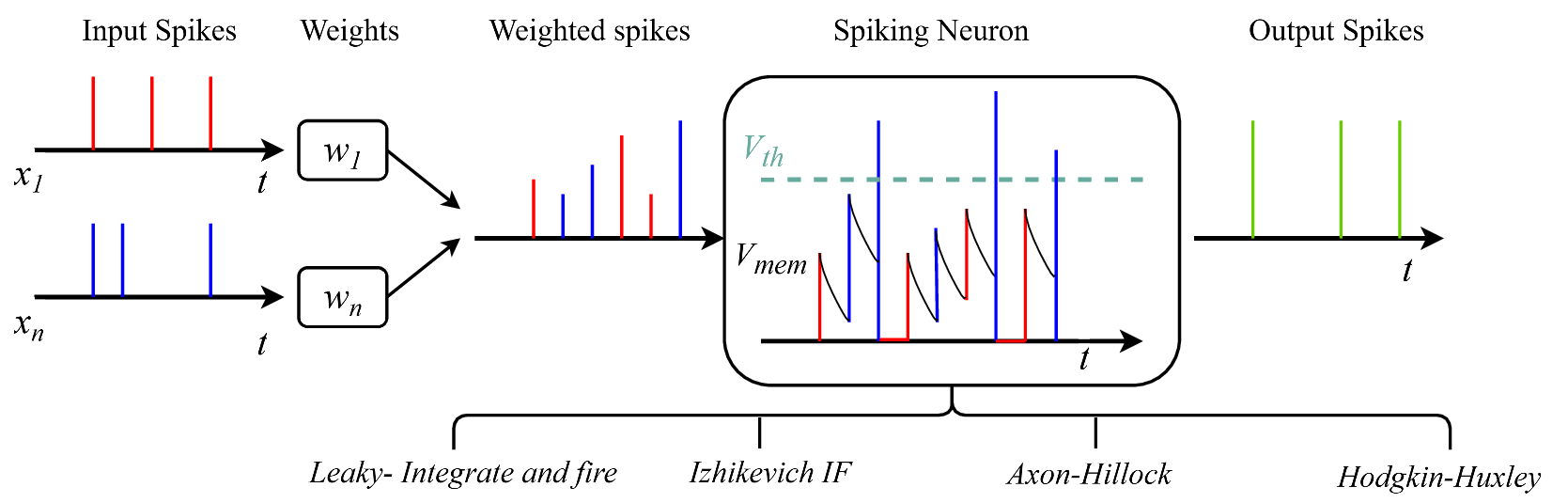, width=\columnwidth}
\end{center}
\caption{Dynamics of representative spiking neuron models: Input spikes are weighted and then integrated over time, triggering an output spike when a defined threshold is reached.}
\label{fig03}
\end{figure}
Unlike conventional artificial neural networks that rely on continuous-valued activations, SNNs encode and transmit information through the timing and occurrence of spikes, enabling inherently sparse and event-driven computation.

From a system perspective, SNNs can be described through two fundamental primitives: \textit{synapses}, which encode and transform incoming spike events through weighted interactions and \textit{neuronal dynamics}, which integrate these inputs and generate output spikes. Fig.~\ref{fig03}, illustrates representative temporal spike behavior arising from these interactions. While high-level models often treat these primitives abstractly, their physical realization in hardware introduces significant complexity.

In practical implementations, particularly in mixed-signal neuromorphic systems, both synaptic and neuronal behaviors are strongly influenced by underlying circuit and device characteristics. Non-idealities such as device mismatch, nonlinear current-voltage relationships, parasitic effects, and process variations can substantially impact signal integration, spike generation, and overall network performance. These effects are especially critical when synaptic weights are stored in analog or non-volatile devices, where physical state variables directly determine computational behavior.

Consequently, accurately modeling SNNs for hardware deployment requires frameworks that capture the interaction between algorithmic dynamics and circuit-level constraints. In this work, we focus on these two key primitives, \textit{analog synapses} and \textit{mixed-signal neuron implementations}, and develop a hardware-aware modeling approach that enables realistic evaluation and design space exploration for neuromorphic systems.
\subsection{Analog Synapses}

In mixed-signal SNNs, synapses are responsible for storing and modulating the weights that govern spike-driven computation. Unlike digital implementations, where weights are abstract numerical values, hardware synapses are realized using physical devices whose electrical characteristics directly determine synaptic behavior.

In this work, the proposed exploration framework incorporates hardware-aware models of analog synapses based on non-volatile devices. Specifically, we support two classes of devices: floating-gate transistors (FG) and resistive random-access memory (ReRAM), which enable persistent storage of synaptic weights while providing tunable conductance for spike-driven current generation.

The following subsections describe the measurement results, device modeling and integration of these analog synapses within the proposed simulation framework.
  
\subsubsection{Floating-Gate based Synapses}~\\
Floating gate transistors (FG) are standard CMOS compatible devices that enable the non-volatile storage of charge, a concept first pioneered by Kahng and Sze in 1967 \cite{kahng1967floating}. Since then, they have been used widely for storing the weights of synapses \cite{diorio_floating-gate_1997,ramakrishnan_floating_2011,liu_temporally_2008,bayat_implementation_2018}. The design of a standard CMOS floating gate device is achieved via the integration of a MOS capacitor to establish a floating node, depicted in the inset of Fig. \ref{fig04}.

By accounting for the charge trapped on the node, the synaptic current $I_\mathrm{synapse}$ modeled using a modified EKV (Enz-Krummenacher-Vittoz) model \cite{enz_analytical_1995,natarajan_modeling_2017} can be given by:

\begin{figure}[htbp]
  \begin{center}
  \epsfig{file=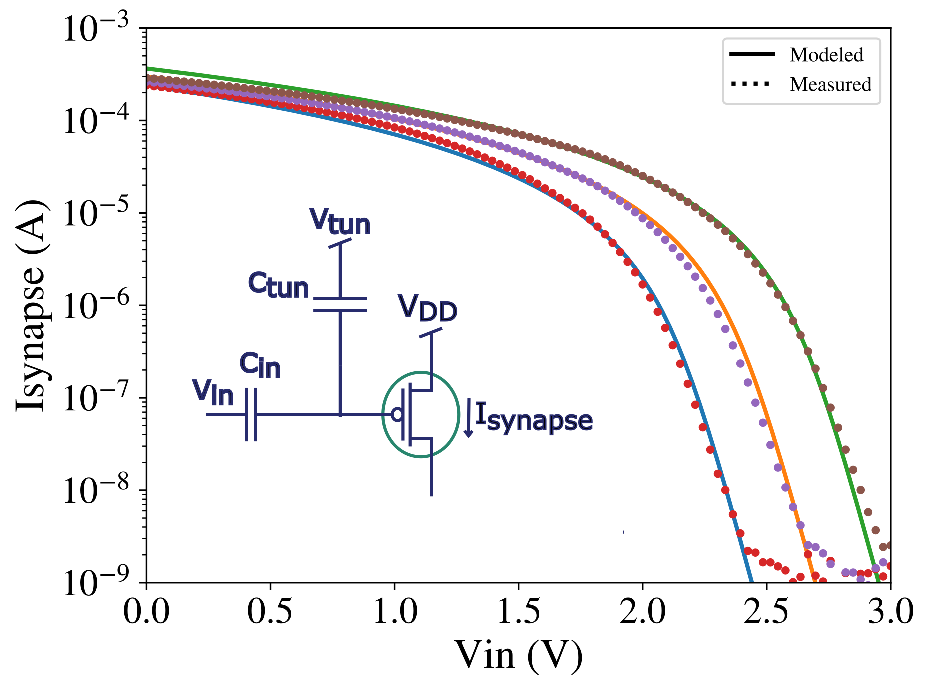, width=0.78\columnwidth}
  \end{center}
  \caption{The figure shows the current vs input voltage relationships of an FG transistor. Measurement results are obtained from FG transistors integrated on a 65 nm CMOS process. Modeling is based on the equation \ref{Eq:Eq1}. Inset shows a transistor level schematic of a direct FG transistor. }\label{fig04}
  \end{figure}

\begin{equation}\label{Eq:Eq1}
\begin{aligned}
I_{\mathrm{synapse}} = I_{\mathrm{thpmos}}
\Bigg[
    \ln^{2}\!\left(
        1 + e^{\frac{\kappa(V_\mathrm{DD} - V_\mathrm{FG} - V_\mathrm{TP}) + \sigma(V_\mathrm{DD} - V_\mathrm{d})}{2U_T}}
    \right)
    - \\
    \ln^{2}\!\left(
        1 + e^{\frac{\kappa(V_\mathrm{DD} - V_\mathrm{FG} - V_\mathrm{TP}) - (V_\mathrm{DD} - V_\mathrm{d})}{2U_T}}
    \right)
\Bigg]
\end{aligned}
\end{equation}

Here, $V_\mathrm{FG}$ is the total floating-gate voltage that depends on:

\begin{equation}\label{Eq:Eq2}
\begin{aligned}
V_\mathrm{FG} \propto 
V_\mathrm{FG_0} 
+ \frac{C_\mathrm{in}}{C_\mathrm{T}} V_\mathrm{in} 
+ \frac{C_\mathrm{gdo}}{C_\mathrm{T}} V_{d}
+ \frac{C_\mathrm{gso}}{C_\mathrm{T}} V_{s}
+ \frac{C_\mathrm{tun}}{C_\mathrm{T}} V_\mathrm{tun}
\end{aligned}
\end{equation}

where $Q_\mathrm{FG}$ denotes the floating-gate charge, $V_\mathrm{in}$ is the input voltage, $V_\mathrm{d}$ is the drain voltage, $V_\mathrm{s}$ is the source voltage (equal to $V_\mathrm{DD}$), and $V_\mathrm{tun}$ is the tunneling voltage. The parameter $\kappa$ represents the floating-gate to channel-surface coupling coefficient, $U_T$ is the thermal voltage, and $\sigma$ models the drain-induced barrier lowering (DIBL) effect. Furthermore, $I_\mathrm{th_{pmos}}$ is proportional to the carrier mobility ($\mu_p$) and the transistor geometry ratio ($W/L$).

In equation (\ref{Eq:Eq2}), $C_\mathrm{in}$ represents the input capacitance, while $C_\mathrm{T}$ denotes the total capacitance associated with the floating-gate node. Specifically, $C_\mathrm{T}$ is the sum of the input capacitance, tunneling capacitance, oxide capacitance and the source/drain overlap capacitances.


As depicted in Fig. \ref{fig04} the input and tunneling capacitances are realized using Metal-Oxide-Semiconductor (MOS) capacitors. Equation \ref{Eq:Eq1} is fitted to the experimental data to extract the specific device parameters required for modeling. Subsequently, the proposed Python-based exploration tool utilizes these fitted parameters for its simulations

\begin{figure}[htbp]
  \begin{center}
  \epsfig{file=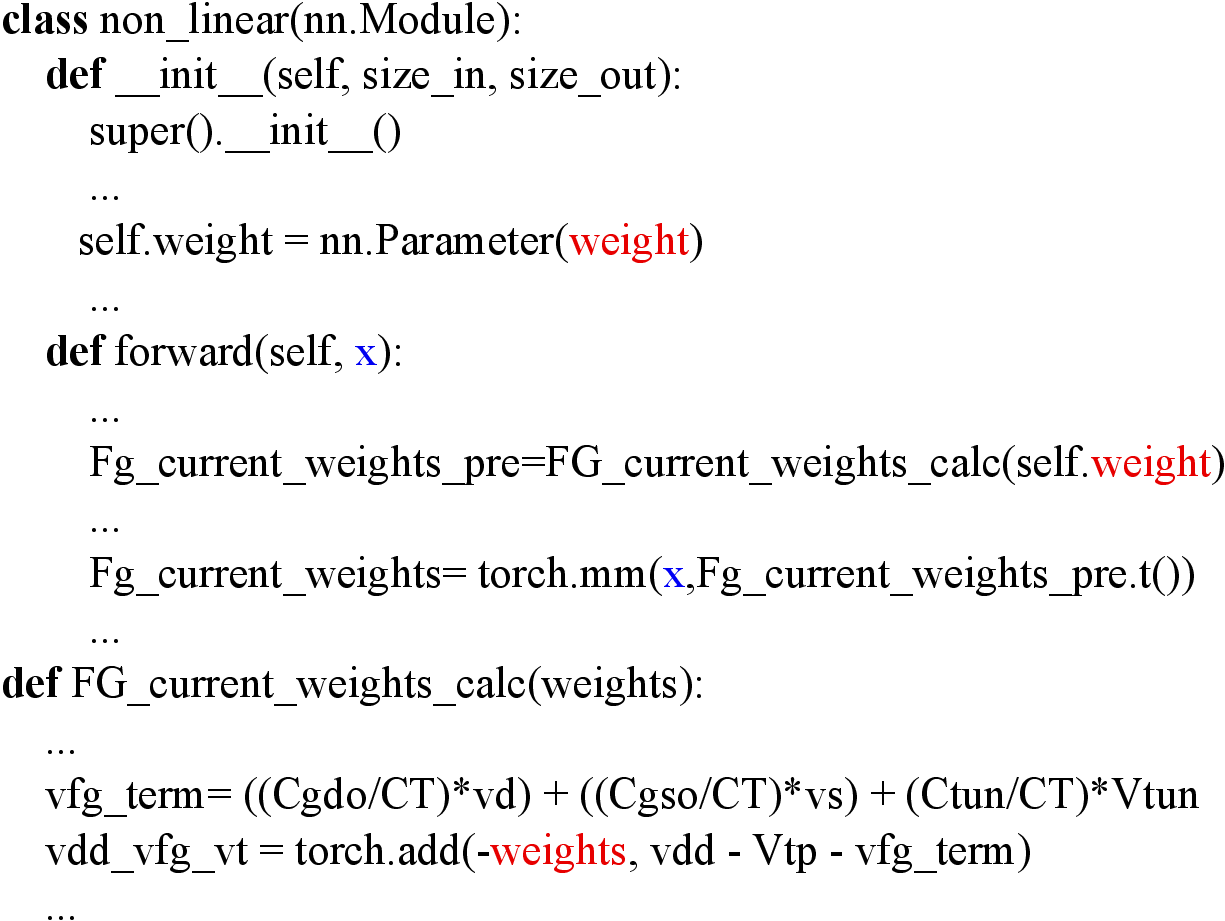, width=0.7\columnwidth}
  \end{center}
  \caption{Figure illustrating the python code for defining a custom pytorch class and torch function to implement FG equation. Here the weight, highlighted in red is $V_\mathrm{FG_{0}}$ from equation \ref{Eq:Eq2}. Further, given the spiking nature of the network we simply multiply the input, highlighted in blue, with the calculated FG current.}\label{fig05}
  \end{figure}
  
 Furthermore, the proposed framework enables direct optimization of the parameter ($V_\mathrm{FG_{0}}$) in equation (\ref{Eq:Eq2}) through a custom PyTorch-based implementation for synaptic current computation. A dedicated module models the FG behavior, where the trainable weight parameter corresponds to ($V_\mathrm{FG_{0}}$), representing the stored charge on the floating gate.

To maintain consistency with the event-driven nature of the network, the input spike train is used to multiplicatively modulate the floating-gate current, yielding the synaptic output current. This approach effectively emulates a pulsed, event-driven synaptic operation of hardware. The architecture of the implemented class and its associated current computation logic are detailed in Fig.~\ref{fig05}.

Since the FG synapse encodes synaptic weights as floating gate charge that modulates the PMOS current, the continuous floating-point weights obtained after training must be mapped to discrete levels that are physically realizable on the FG device. To achieve this, an 8-bit fixed-point quantization scheme is applied to the trained weights. Our and other previous work has shown that FG transistors can provide 8-bits of accuracy \cite{chowdhury2024analysis,kim_integrated_2016}.

The trained weights ($V_\mathrm{FG_{0}}$) are mapped through the FG current model and projected onto the nearest realizable quantization level (pertaining to 8-bit quantization) within the valid current range supported by the FG device determined from measured device characteristic. Weights exceeding the representable range are clipped to the nearest boundary level to avoid overflow and ensure hardware compatibility. Post-quantization inference is evaluated to verify that classification accuracy is maintained despite the reduced precision, confirming the robustness of the FG synapse-based SNN to weight discretization.

\subsubsection{ReRAM Analog Synapses}~\\
A second set of analog synapses that tool currently supports is a Resistive Random Access Memory (ReRAM). ReRAMs are two-terminal devices capable of being programmed to different resistance states. Their structure consists of a metal oxide layer sandwiched between two conducting electrodes \cite{brando2023modeling}. Various oxide switching materials, including HfOx, TaOx and TiOx \cite{govoreanu201110,sokolov2018influence,jeong2010low} deposited via atomic layer deposition, have demonstrated switching capabilities. 
\begin{figure}[htbp]
  \begin{center}
  \epsfig{file=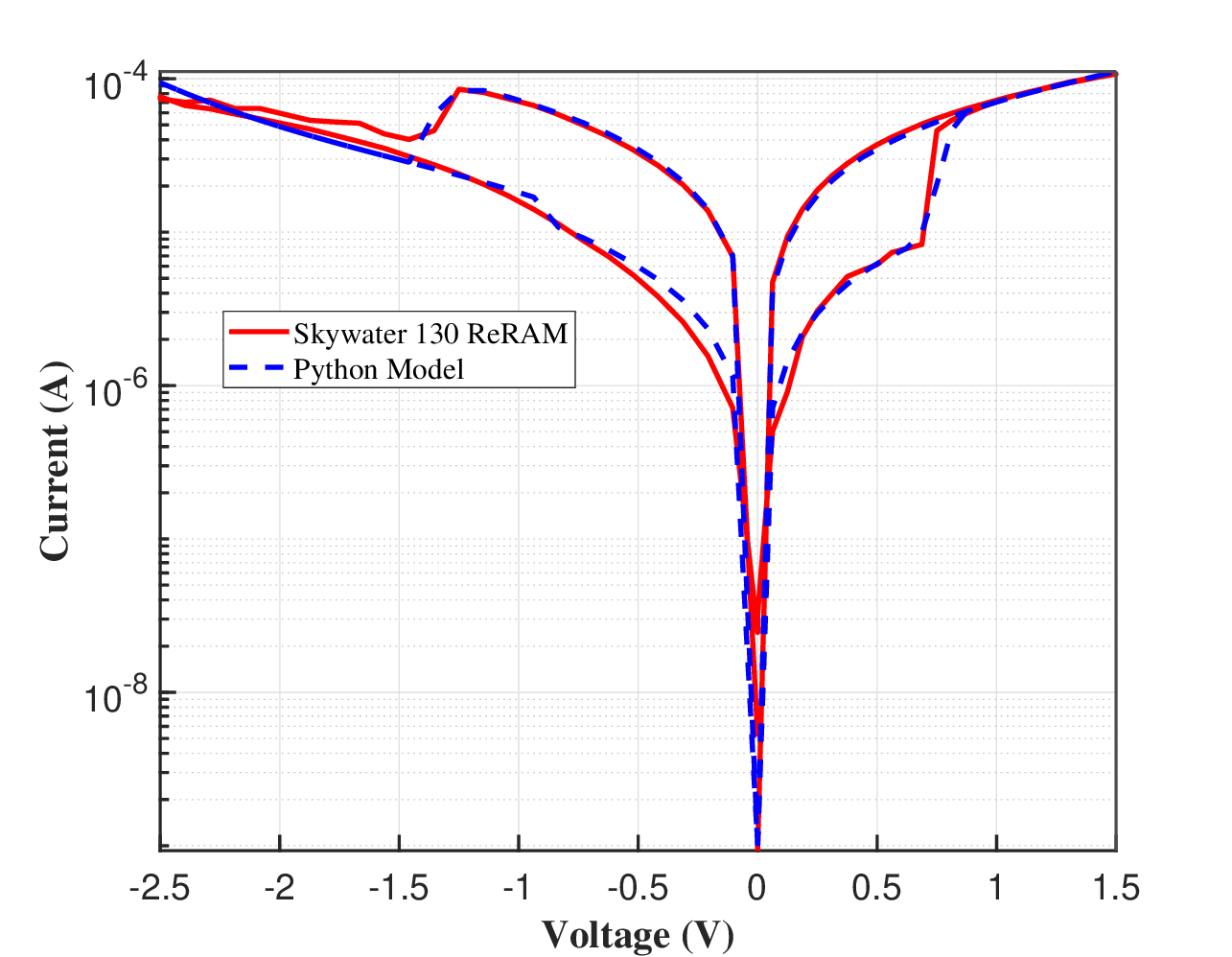, width=0.6\columnwidth}
  \end{center}
  \caption{Current-voltage (I-V) characteristics of the ReRAM device. The ReRAM devices is characterized extensively in the following work \cite{didin2026characterization}}\label{fig06}
  \end{figure}
In this work, we use ReRAM implemented in Skywater Technologies' 130 nm CMOS process fabricated through Front-End-Of-Line (FEOL) processing, where the device replaces the via between Metal 1 and Metal 2. This configuration places an HfOx switching oxide between a TiN-Ti top electrode and a TiN bottom electrode.

Fig. \ref{fig06} shows the current vs. voltage characteristics of the ReRAM. Applying a positive voltage across the device (set) decreases its resistance. While applying a negative voltage across the device (reset) will increase its resistance\cite{didin2026characterization}. These transitions are associated with the distribution of oxygen vacancies controllable through set (positive) and reset (negative) pulses. During inference a voltage lower than set pulses can be used to measure the device without changing the resistance.

However, to accurately map weights trained using high-level language such as python requires embedding the non-linear model of the ReRAM as part of the learning process. The I-V characteristics of the device is non-linear given by the equation \ref{equ:reramiv}.

\begin{equation}
    I(V) = I_\mathrm{0}\exp{\bigg(-\frac{g}{g_\mathrm{0}}\bigg)}\sinh{\bigg(\frac{V}{V_\mathrm{0}}\bigg)}
    \label{equ:reramiv}
\end{equation}

where $I$ denotes the device current, $I_\mathrm{0}$, $g_0$, and $V_0$ are calibration parameters derived from physical device measurements and $g$ represents the gap distance between the filament tip and the opposing electrode. The framework enables direct optimization of $g$ in equation \ref{equ:reramiv} by implementing a custom PyTorch class for synaptic current calculation. In this work, the device model is designed to capture the mean device behavior. (While physical implementations are subject to device-to-device and cycle-to-cycle conductance variations).

\begin{figure}[h]
  \begin{center}
  \epsfig{file=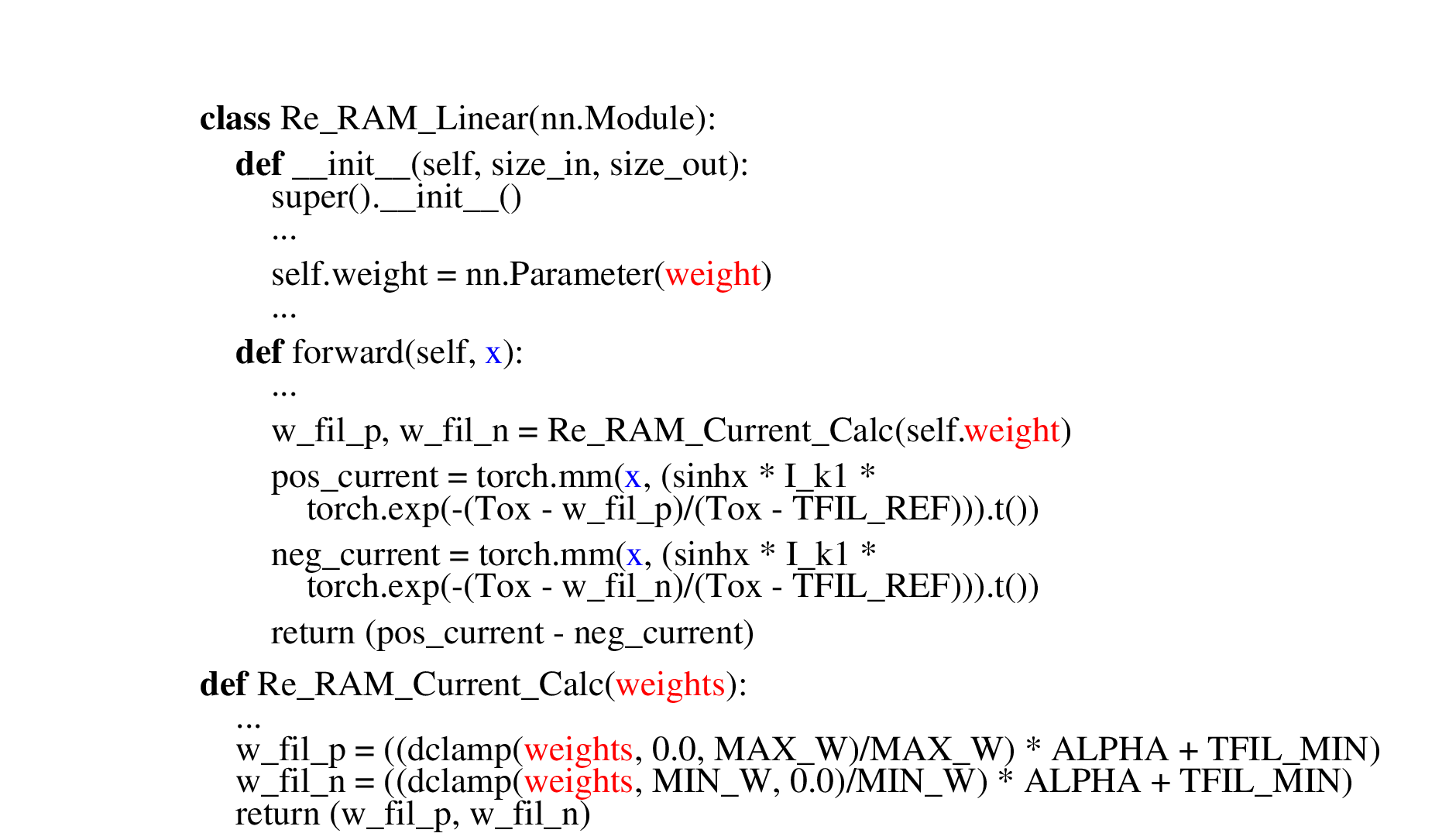, width=0.9\columnwidth}
  \end{center}
  \caption{ A demonstration of Python code that defines a custom PyTorch class and torch function for computing ReRAM current. The weight parameter (highlighted in red) corresponds to $g$ in equation \ref{equ:reramiv}. Given the spiking nature of the network, the input (highlighted in blue) is directly multiplied by the calculated ReRAM current.}\label{fig07}
  \end{figure}

Fig. \ref{fig07} shows the Python implementation of the custom PyTorch class used to compute the ReRAM synaptic current. In this implementation, the trainable weight parameter corresponds to $g$  in equation (\ref{equ:reramiv}). Due to the spiking nature of the network, the input signal is multiplicatively combined with the computed ReRAM current to produce the synaptic output.

To further constrain the ReRAM weights to discrete programmable conductance levels, $3$-bit quantization scheme was also applied post-training for the ReRAM synapses, where the clamped weight values are discretized into a fixed number of quantization levels within the filament thickness range ($T_{fill,min}$, $T_{fill,max}$). Quantization is performed by a nearest-neighbor projection onto 8 discrete levels derived from experimentally measured ReRAM conductance states ~\cite{didin2026characterization}. This ensures that each synaptic weight maps to a physically distinct and programmable resistance state of the ReRAM device. 

\subsection{Neuron Models}

Biological neurons can be modeled at varying levels of abstraction when designing SNNs or implementing neuromorphic hardware in silicon. Common neuron models such as Axon-Hillock (AH), Leaky Integrate-and-Fire (LIF), Izhikevich, and Hodgkin–Huxley (HH) offer different trade-offs between biological fidelity, computational complexity, and hardware efficiency~\cite{indiveri_neuromorphic_2011}. 

In this work, the proposed framework incorporates multiple neuron models, including AH, LIF and HH, enabling comparative evaluation across different levels of modeling accuracy. The modular design of the tool also allows for straightforward integration of additional neuron models, facilitating flexible exploration of neuron-level design choices in mixed-signal neuromorphic systems.

\subsubsection{Axon-Hillock Neuron}
\begin{figure}[h]
    \centering

    \begin{subfigure}{0.28\textwidth}
        \centering
        \epsfig{file=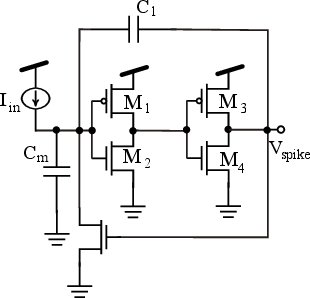, width=\columnwidth}
        \caption{}
      
    \end{subfigure}
    \hfill
    \begin{subfigure}{0.65\textwidth}
        \centering
        \epsfig{file=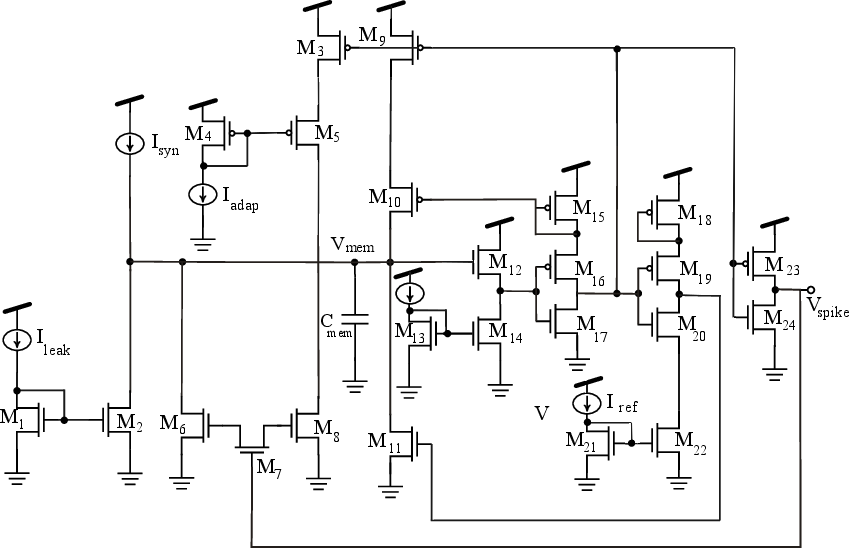, width=\columnwidth}
      \caption{}
      
    \end{subfigure}

    \caption{(a) Axon hillock circuit (b) LIF neuron circuit}
    \label{fig08}
\end{figure}

~\\
Axon-Hillock circuit is a simplistic neuron model that captures the integrate and fire capability of a neuron \cite{mead_analog_1989}. Input synaptic current charges the membrane capacitor till the voltage reaches a specific threshold. Based on the sizing of the transistors in the inverters, seen in Fig. \ref{fig08}(a), the circuit will emit a spike at a specific threshold voltage (500 mV-700 mV depending on the size and threshold voltages of NMOS and PMOS transistors). The capacitor ($C_\mathrm1$) acts as a positive feedback (mimicking the sodium channel in biological neuron) which pulls the membrane potential further up. The output spike also triggers the reset mechanism in the circuit by leaking some amount of charge to the ground (based on the biasing of transistor M5). In a biological neuron the potassium channels perform the reset. The following equation is used to simulate the behaviour of charging of the membrane capacitance and the effect of feedback capacitor:
\begin{equation}\label{Eq:Eq3}
C_{\mathrm{m}} \frac{dV_{\mathrm{mem}}}{dt}
= I_{\mathrm{in}}
+ C_{\mathrm{1}} \frac{d}{dt}\!\left(V_{\mathrm{out}} - V_{\mathrm{mem}}\right)
\end{equation}

Here, $I_{\mathrm{in}}$ is the synaptic input current.

Simple forward Euler update is followed, discretised at each timestep:

\begin{equation}
    V_\mathrm{new} = V_\mathrm{mem} + \frac{dV_\mathrm{mem}}{dt} \cdot \Delta t
\end{equation}
Upon spike emission, the membrane potential is reset to with a surrogate replacing the non-differentiable Heaviside to enable gradient flow through the threshold crossing during training:
\begin{equation}
V_\mathrm{out} = \Theta(V_{new} - V_{th}) = 
\begin{cases} 
1 & \text{if } V_{new} \geq V_{th} \\ 
0 & \text{otherwise} 
\end{cases}, 
\qquad 
V \leftarrow (1 - s_{out}) \cdot V_{new}
\end{equation}
 Fig. \ref{fig09} shows the comparison of Inter-Spike-Interval (ISI) between transistor level circuit simulation in 65 nm CMOS node and the output obtained from the tool.

\subsubsection{Leaky-Integrate and Fire Neurons}~\\
The AH circuit has several drawbacks including high power consumption, spiking threshold that is fixed, and lacks spike frequency adaptation. An adaptive Leaky-Integrate-and-Fire (LIF) neuron introduced in \cite{1206342} provides an alternative design enabling spike frequency adaptation, adaptive threshold and lower-power consumption. This tool models this version of the adaptive LIF as depicted in Fig. \ref{fig08}(b). 
The membrane potential dynamics of the LIF neuron in the tool are governed by:

\begin{equation}
    \frac{dV_\mathrm{mem}}{dt} = \frac{I_\mathrm{syn}-I_\mathrm{leak} - I_\mathrm{adapt} - I_\mathrm{ref} \cdot R }{C_\mathrm{mem}}
\end{equation}

where, $I_\mathrm{syn}$ is the synaptic input current, $I_\mathrm{leak}$ is a constant leak current that passively drives the membrane 
potential toward rest, $I_\mathrm{adapt}$ is the adaptive current that models spike-frequency adaptation, $I_\mathrm{ref} \cdot R$ 
is the refractory suppression term and $C_\mathrm{mem}$ is the membrane capacitance.

The membrane potential is updated at each discrete timestep $\Delta t$ using 
first-order Euler integration:

\begin{equation}
    V_\mathrm{mem}^{t+1} = V_\mathrm{mem}^{t} + \frac{dV^t_\mathrm{mem}}{dt} \cdot \Delta t
\end{equation}

A spike is generated when the membrane potential exceeds the threshold $V_{th}$, 
approximated using a surrogate gradient function $\sigma(\cdot)$ to enable 
backpropagation through the discontinuous threshold operation:

\begin{equation}
    S = \sigma(V_\mathrm{mem}^{t+1} - V_\mathrm{th}) \approx \Theta(V_\mathrm{mem}^{t+1} - V_\mathrm{th})
\end{equation}
Following a spike, a reset is applied to the membrane potential, 
zeroing only the neurons that fired while preserving the state of inactive neurons:
\begin{equation}
    V_\mathrm{reset} \leftarrow (1 - S) \cdot V_\mathrm{mem}^{t+1}
\end{equation}
The refractory state variable $R$ rises upon spiking and decays exponentially 
with time constant $\tau_{ref}$, suppressing re-firing in the immediate 
post-spike period. The adaptive current $I_\mathrm{adapt}$ is set to a fixed penalty value upon each spike 
and retains its previous value otherwise
\begin{equation}
    I_\mathrm{adapt} = S \cdot I_\mathrm{adapt,const} + (1 - S) \cdot I_\mathrm{adapt}^{t}
\end{equation}

Between spikes, $I_\mathrm{adapt}$ decays exponentially with time constant $\tau_\mathrm{adapt}$, 
with the $(1-S)$ gate ensuring decay occurs only in the absence of a spike.
 These parameters are matched with circuit simulations performed in cadence for 65 nm CMOS node and the output spike rate of the model and transistors level simulation are compared. 
\begin{figure}[htbp]
  \begin{center}
  \epsfig{file=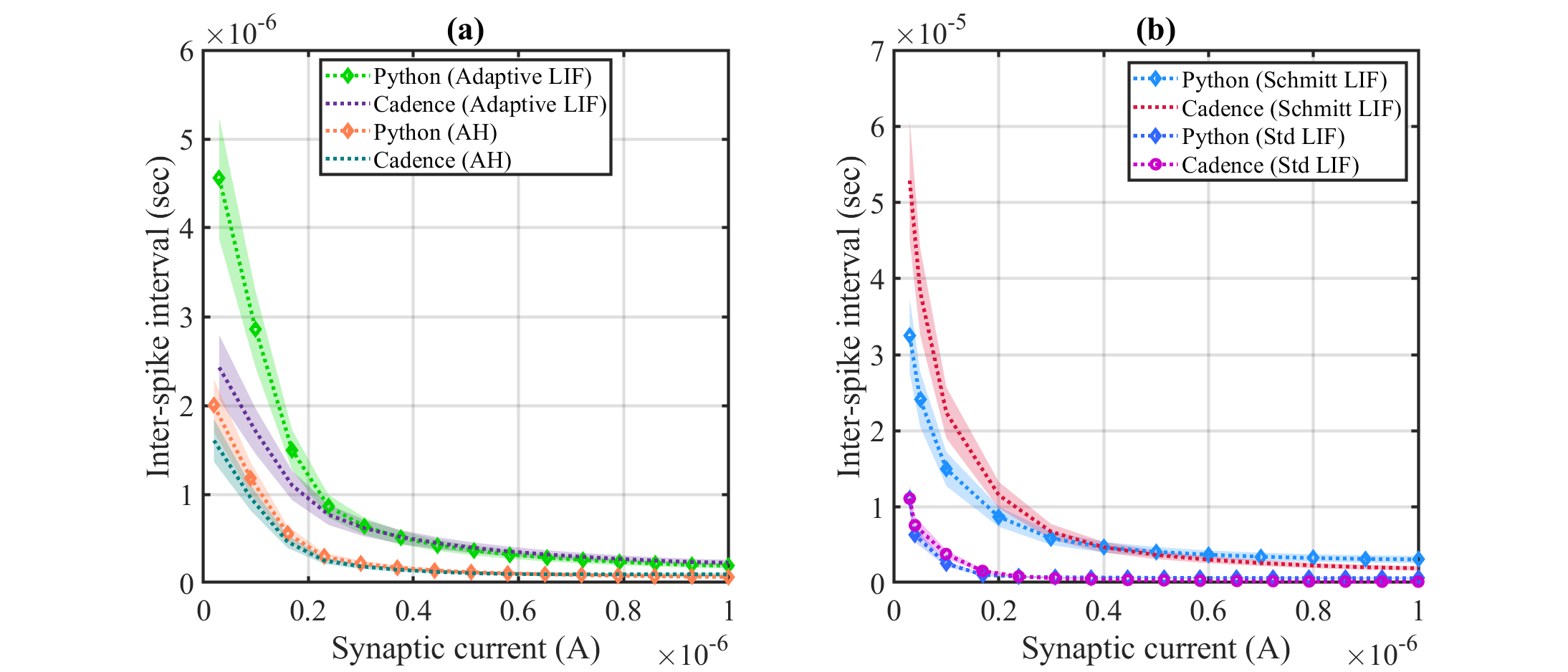, width=0.9\columnwidth}
  \end{center}
  \caption{Comparison of ISI vs. synaptic current between the Python model and transistor-level circuit implementation of  analog (a) Adaptive LIF and Axon-Hillock neurons in 65nm technology node (b) Schmitt LIF neurons and Standard LIF in 28 nm technology node
  }\label{fig09}
  \vspace{-5mm}
\end{figure}
Fig. \ref{fig09} compares ISI for the adaptive LIF neuron in cadence and python for a variety of synaptic input currents. The analog circuit used for adaptive LIF neurons is similar to the one presented in \cite{1206342}\cite{chowdhury_hardware_2022}. 

Two additional neuron models, namely the Standard LIF and Schmitt-triggered LIF neurons proposed in \cite{11412426}, are also incorporated into the framework to enable broader comparative analysis of circuit-level neural dynamics. The parameters for these models are extracted from a 28\,nm CMOS process, demonstrating the flexibility of the proposed tool to support neuron implementations across multiple technology nodes and PDKs. 

The Standard LIF neuron employs a membrane capacitor to integrate synaptic input currents, with leakage behavior modeled using a current mirror and spike generation realized through a CMOS inverter-based threshold comparator followed by a capacitor reset mechanism. In contrast, the Schmitt-triggered LIF neuron utilizes the hysteresis characteristics of a Schmitt trigger to establish distinct firing and reset thresholds, providing improved noise immunity and more controlled spike widths compared to the Standard LIF implementation. The circuit schematics, the temporal deviation between the membrane potential and spike output obtained from the Python model, and that from the Cadence simulation are presented in the supplementary material for all neuron models.

\subsubsection{Hodgkin-Huxley Neuron}~\\
Another neuron model explored in this tool is the Hodgkin-Huxley (HH) model, developed by Alan Hodgkin and Andrew Huxley in 1952 \cite{HH}, a mathematical framework that describes how action potentials are initiated and propagated in neurons. Based on their experiments with the giant squid axon, the model represents the cell membrane as an electrical circuit, with ion channels for sodium ($Na^+$) and potassium ($K^+$) acting as variable conductances. The action potential arises from the coordinated activity of voltage-gated ion channels, whose exponential carrier distributions and ion transport mechanisms drift and diffusion across a bilipid membrane closely parallel charge transport through a transistor channel. The key insight driving silicon implementations is that ion flow through biological channels and electron flow through MOSFET transistors share the same fundamental driving forces: drift and diffusion, both producing an exponential current-voltage relationship. 

\begin{figure}[htbp]
    \centering
    \includegraphics[width=0.88\columnwidth]{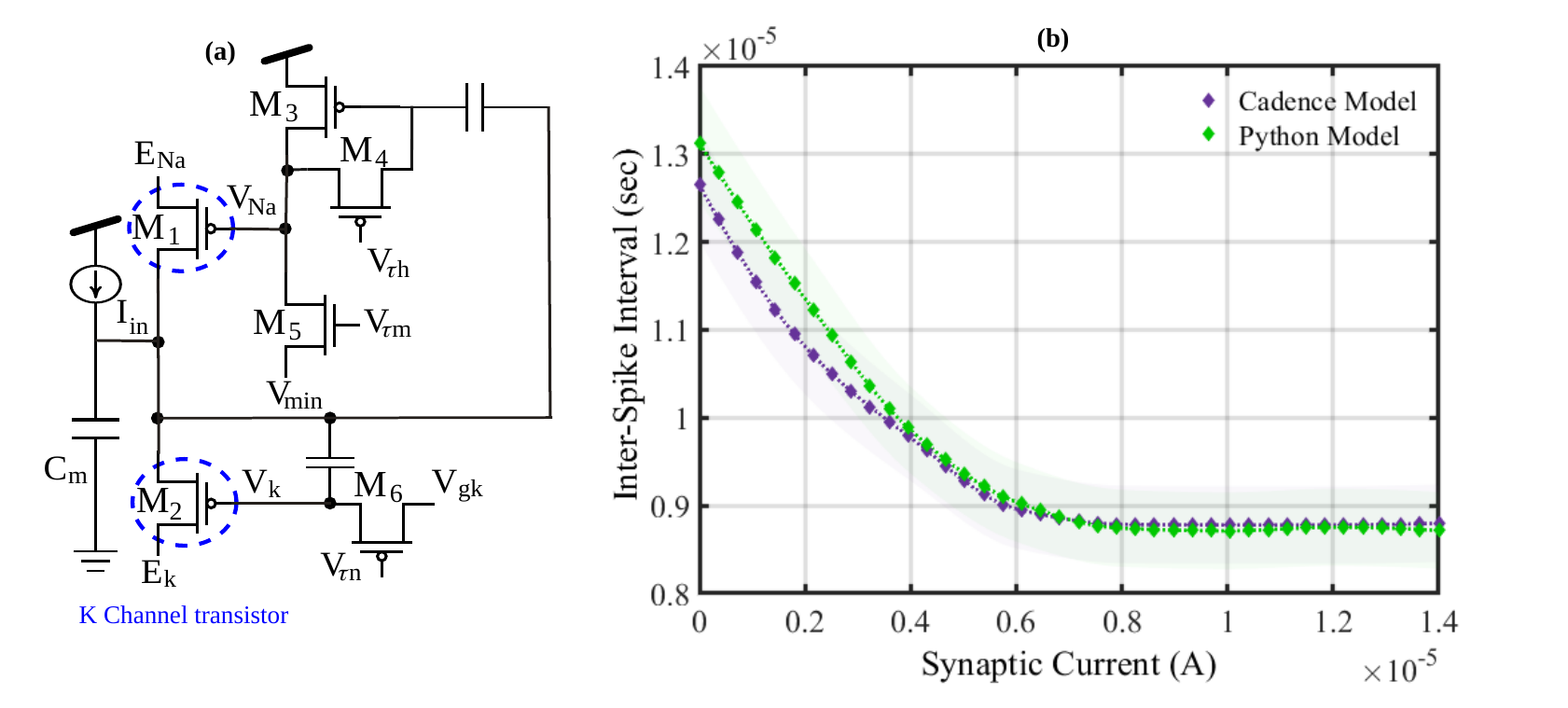}
    \caption{(a) Hodgkin-Huxley neuron model (b) ISI vs. synaptic current for HH model (65 nm process implementation versus python model)}
    \label{fig10}
\end{figure}

Exploiting this analogy, the $Na^{+}$ channel is implemented using a subthreshold MOSFET configured as a bandpass filter \cite{natarajan_hodgkinhuxley_2018} capturing the channel's fast activation and slower inactivation dynamics through two tunable time constants ($\tau_m$ and $\tau_h$) as shown in Fig.\ref{fig10}(a). The $K^{+}$ channel, which only activates (no inactivation) and responds more slowly, is implemented as a low-pass filter using a single transistor whose nonlinear conductance naturally reproduces the characteristic S-shaped current response seen in biology. Together, these two transistor-based channel circuits interact on a shared membrane capacitance node to generate action potentials that closely match biological recordings in shape, magnitude and timing. 
The total membrane current ($Iin$) is the sum of the capacitive and ionic currents, defined by:$$I_{in} = C_\mathrm{m} \frac{dV_\mathrm{m}}{dt} + g_\mathrm{K}(V_\mathrm{m} - V_\mathrm{K}) + g_\mathrm{Na}(V_\mathrm{m} - V_\mathrm{Na}) + g_\mathrm{l}(V_\mathrm{m} - V_\mathrm{l})$$
where: $C_\mathrm{m}$ is the membrane capacitance per unit area, $g_\mathrm{Na}, g_\mathrm{K}, g_\mathrm{l}$ are conductances for sodium, potassium and leakage; and $V_\mathrm{Na}, V_\mathrm{K}, V_\mathrm{l}$ are the respective reversal potentials.

The HH model was implented using 65 nm node and the ISI for different  synaptic input currents for the simulated and Python neuron model is as depicted in Fig.\ref{fig10}(b). The temporal deviation between the membrane potential obtained from the Python model and that from the Cadence simulation is presented in the supplementary material.

\section{Architecture Implementation}

Building upon the synaptic and neuronal primitives described in the previous sections, we construct network-level SNN architectures that enable system-level evaluation of mixed-signal neuromorphic designs. In these networks, spike-based information propagates through layers of neurons interconnected by hardware-realistic synapses, where synaptic weights are implemented using non-volatile devices.

\begin{figure}[htbp]
    \centering
    \includegraphics[scale=0.65]{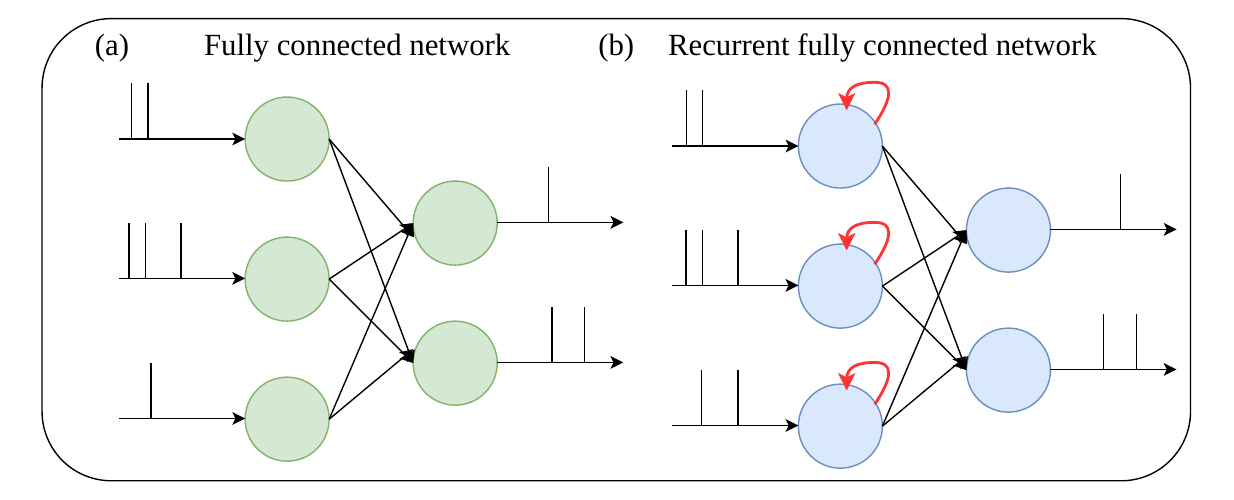}
    \caption{(a) Fully connected architecture (b) Recurrent connected architecture}
    \label{fig11}
\end{figure}

The proposed framework supports both fully connected and recurrent architectures, enabling flexible exploration of spatial and temporal processing capabilities. Fully connected architectures provide dense feedforward connectivity for hierarchical feature extraction, while recurrent architectures introduce feedback connections that enable temporal dynamics and memory. The corresponding connectivity patterns are illustrated in Fig.~\ref{fig11}.

In the fully connected SNN (FC-SNN), each neuron receives inputs from all neurons in the preceding layer through weighted synaptic connections. These synapses are implemented using FG or ReRAM devices, allowing direct mapping of synaptic weights to physical device states. To capture the excitatory and inhibitory nature of synaptic interactions, each connection is realized using differential pairs of synapses, where one branch contributes positive (charging) current and the other contributes negative (discharging) current to the neuron state. This configuration enables representation of both weight polarities while maintaining compatibility with hardware constraints.

The recurrent SNN (R-SNN) extends the feedforward architecture by incorporating temporal feedback, enabling the network to capture dependencies across time. In this configuration, a recurrent layer of spiking neurons receives both external inputs and feedback signals through hardware-aware synapses implemented using FG or ReRAM devices. The internal state of the neurons is therefore influenced by both current inputs and past activity, allowing the network to encode temporal context.

This recurrent layer is followed by a fully connected spiking layer, forming a hybrid architecture that combines temporal memory with hierarchical processing. Both layers interface with a readout stage that computes the network output and facilitates gradient-based optimization of synaptic parameters discussed in detail in the next section. By explicitly modeling synaptic weights as device-level quantities (e.g., floating-gate voltage or ReRAM conductance), the framework enables consistent training and evaluation across both architectures under realistic hardware constraints.

\section{Learning Algorithms and Hardware-Aware Training Methodology}

Building upon the hardware-aware synaptic and neuronal models described in the previous sections, the proposed framework employs a global learning strategy that directly optimizes device-level synaptic parameters. Unlike conventional approaches that treat synaptic weights as abstract numerical values, in this work each synaptic weight $W_{ij}$ is mapped to a physical device parameter $\theta_\mathrm{ij}$, such as the floating-gate voltage ($V_\mathrm{FG_0}$) for FG synapses or the filament gap ($g$) for ReRAM devices. Thus, learning is performed directly in the device domain, ensuring consistency between training and hardware implementation.

To enable effective credit assignment across time, we adopt backpropagation through time (BPTT)~\cite{lillicrap2019backpropagation}, which computes a global loss at the network output and propagates gradients backward through all layers and time steps. In SNNs, BPTT is adapted using surrogate gradient methods \cite{neftci_surrogate_2019} to handle the non-differentiable spike generation function.

A single readout layer computes the global loss $\mathcal{L}$, and gradients are propagated through the unrolled network over $T$ time steps. The gradient of the loss with respect to the synaptic weights is expressed as:

\begin{equation}
\frac{\partial \mathcal{L}}{\partial W_{ij}} = \sum_{t=1}^{T} \frac{\partial \mathcal{L}}{\partial U_i^t} \cdot \frac{\partial U_i^t}{\partial W_{ij}}
\end{equation}

where the dependence of membrane potential on synaptic input is given by:

\begin{equation}
\frac{\partial U_i^t}{\partial W_{ij}} = S_j^{t-1}
\end{equation}

Here, $\mathcal{L}$ denotes the global loss, $W_{ij}$ is the effective synaptic weight from neuron $j$ to neuron $i$, $U_i^t$ is the membrane potential at time step $t$, and $S_j^{t-1}$ is the presynaptic spike.

Since the weights are functions of device parameters ($W_{ij} = f(\theta_{ij})$), the learning rule is applied through the chain rule:

\begin{equation}
\frac{\partial \mathcal{L}}{\partial \theta_{ij}} = 
\frac{\partial \mathcal{L}}{\partial W_{ij}} \cdot 
\frac{\partial W_{ij}}{\partial \theta_{ij}}
\end{equation}

where $\theta_{ij} \in \{V_{FG_0}, g\}$ depending on the synapse type. This formulation enables direct updates to physical device parameters during training.

The temporal error signal $\delta_i^t = \frac{\partial \mathcal{L}}{\partial U_i^t}$ is computed as:

\begin{equation}
\delta_i^t = \frac{\partial \mathcal{L}}{\partial S_i^t} \cdot \frac{\partial S_i^t}{\partial U_i^t} + \beta \cdot \delta_i^{t+1}
\end{equation}

where $\frac{\partial S_i^t}{\partial U_i^t}$ is approximated using a surrogate gradient, and $\beta$ is the membrane decay factor that propagates error across time.

By integrating surrogate gradient-based BPTT with device-level parameter optimization, the proposed framework enables end-to-end training of mixed-signal SNNs. This approach ensures that learning dynamics inherently account for device non-idealities, eliminating the need for post-training weight mapping and enabling accurate prediction of hardware performance.

\subsection{Spiking Dataset and Pre-Processing}

\subsubsection{Benchmark Datasets}~\\
The proposed framework is evaluated on three widely used neuromorphic benchmark datasets spanning multiple sensing modalities and levels of temporal complexity. These datasets are selected to assess the performance of the hardware-aware SNN models under diverse input characteristics, including vision-based and auditory spike streams.

The \textbf{N-MNIST} (Neuromorphic-MNIST) dataset \cite{orchard_converting_2015} is an event-driven extension of the classical MNIST handwritten digit dataset, generated by recording static images using a DVS128 camera with controlled saccadic motion. It consists of $60{,}000$ training samples and $10{,}000$ test samples at a spatial resolution of $34 \times 34$, across $10$ digit classes.

The \textbf{DVS Gesture} dataset \cite{amir2017low} captures dynamic hand and arm gestures using a DVS128 sensor. It contains $1{,}342$ recordings of $11$ gesture classes performed by $29$ subjects under varying lighting conditions, with a spatial resolution of $128 \times 128$. This dataset is commonly used to evaluate spatio-temporal processing capabilities in event-based vision systems.

The \textbf{Spiking Heidelberg Digits (SHD)} dataset \cite{9311226} represents an auditory benchmark, where spoken digits ($0$--$9$) in English and German are converted into spike-based representations. The dataset consists of $8{,}156$ training samples and $2{,}264$ test samples, encoded as temporal spike trains across $700$ input channels. SHD is particularly useful for evaluating the ability of SNNs to capture temporal dependencies in sequential data.

Together, these datasets enable comprehensive evaluation of the proposed framework across spatial, temporal, and multi-modal spiking inputs, providing insight into the effectiveness of hardware-aware training and device-level synaptic modeling under realistic operating conditions.

\subsubsection{Data Pre-processing}~\\
Event-based neuromorphic datasets such as \textbf{N-MNIST}, \textbf{DVS Gesture}, and \textbf{SHD} produce asynchronous spatio-temporal spike streams with high event density, which can be computationally expensive for training and deployment on neuromorphic hardware. Preprocessing is therefore required to reduce data dimensionality while preserving key spatio-temporal features.

Conventional approaches often rely on frame-based accumulation~\cite{zheng2023deep, lenz2021tonic}, which simplifies training but sacrifices temporal precision and undermines the event-driven nature of SNNs. To address these limitations, we adopt a fully event-driven preprocessing pipeline consisting of spatial pooling and temporal discretization.

\textbf{Spatial Processing:} 
We employ Mutual SNN Pooling~\cite{gruel2023performance}, a biologically inspired method that operates directly on spike events without intermediate frame formation. 
\begin{figure}[htbp]
  \begin{center}
  \epsfig{file=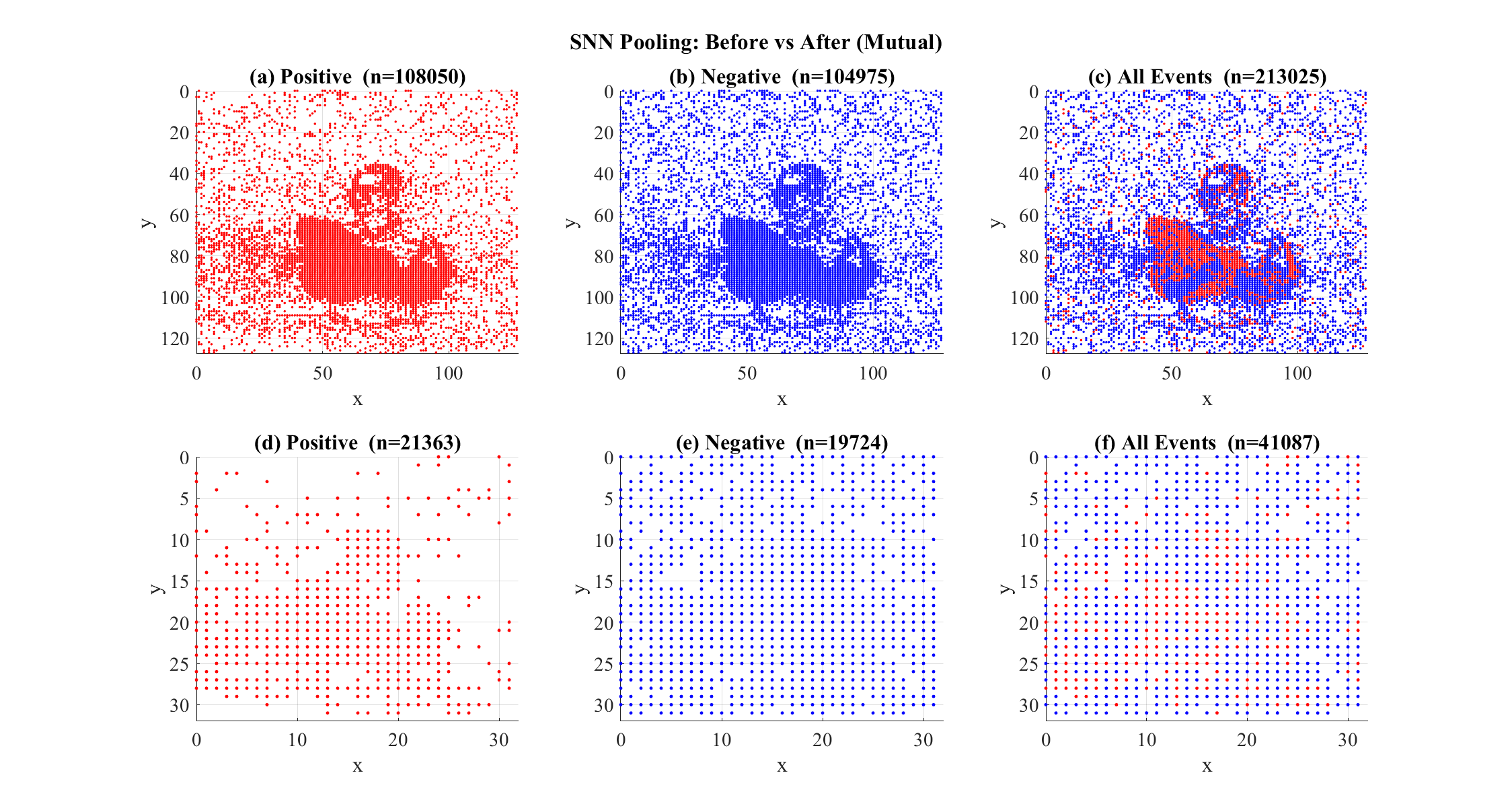, width=\columnwidth}
  \end{center}
  \caption{Visualization of DVS Gesture dataset events spatial downsampling from $128\times128$ to  $32\times32$ resolution, showing positive (red) and negative (blue) polarity events before and after mutual SNN pooling (Here $n$: number of events).
  }\label{fig12}
  \vspace{-5mm}
  \end{figure}
  
This approach uses competitive dynamics between excitatory and inhibitory neurons to retain salient events while suppressing redundant activity. For the DVS Gesture dataset, this reduces the spatial resolution from $128 \times 128$ to $32 \times 32$, significantly lowering computational cost while preserving spatio-temporal structure (Fig.~\ref{fig12}).

\textbf{Temporal Processing:}  
The event streams are discretized into $T$ time bins, producing a binary spike tensor of size $(T \times C \times H \times W)$. Within each bin, a pixel is assigned a value of 1 if one or more events occur. This representation preserves sparsity while enabling efficient training.

Importantly, binary spike encoding aligns with mixed-signal neuromorphic hardware, where synaptic inputs are processed as current pulses. This avoids large dynamic ranges and ensures compatibility with analog synapse implementations such as floating-gate and ReRAM devices.

The choice of $T$ provides a trade-off between temporal resolution and computational cost, with larger $T$ preserving finer timing information and smaller $T$ improving efficiency. This is particularly important for high-density datasets such as DVS Gesture and SHD.

Overall, the proposed preprocessing pipeline preserves the event-driven characteristics of neuromorphic data while enabling efficient and hardware-compatible SNN training.

\section{Results and Analysis}

To validate the proposed hardware-aware SNN framework, we evaluate its ability to implement the non-linearly separable XOR logic function using mixed-signal components. The results obtained from the proposed tool are compared against SPICE-based transistor-level simulations performed in Cadence. In the Cadence implementation, the network is realized in a 65 nm CMOS process using floating-gate (FG) based synapses and adaptive Leaky Integrate-and-Fire (LIF) neurons.

The adopted architecture is a fully connected 2 $\times$ 8 $\times$ 2 network, consisting of two input neurons, an 8-neuron hidden layer, and two output neurons ($N_1$ and $N_2$). Each synaptic connection is realized using differential FG synapses, comprising one positive and one negative floating-gate device to represent both weight polarities. This configuration enables signed weight representation while maintaining compatibility with analog hardware constraints.

The Cadence implementation utilizes Verilog-A models of floating-gate (FG) devices and is simulated at the transistor level to capture device-level behavior. The resulting design occupies an area of 0.00354\,mm$^2$ and exhibits a power consumption of 24.9\,$\mu$W under nominal operating conditions.
\subsection{SNN XOR Validation}

\begin{figure}[htbp]
  \centering
  \epsfig{file=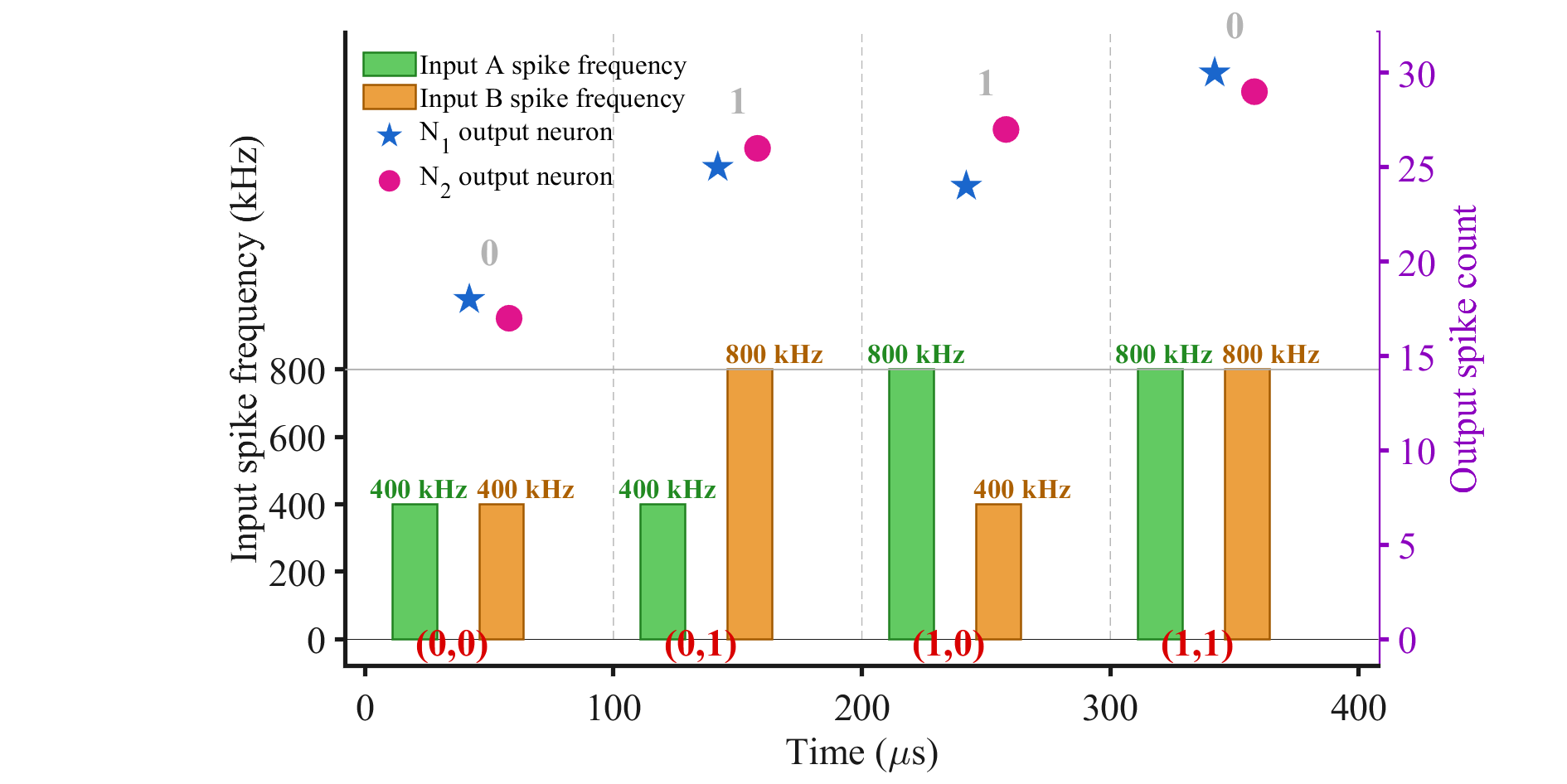, width=\columnwidth}
  \caption{XOR Logic Decoding using Spiking Neural Activity in 65 nm} CMOS process using Adaptive LIF\label{fig13}
\end{figure}
For evaluation, the four XOR input combinations—(A=0, B=0), (A=0, B=1), (A=1, B=0), and (A=1, B=1) are encoded as spike trains with distinct input spike frequencies. Each input pair is applied over a 100\,$\mu$s time window, resulting in a total evaluation duration of 400\,$\mu$s. As shown in Fig.~\ref{fig13}, input spike frequencies of 400\,kHz and 800\,kHz are used to represent binary values, allowing clear separation between logical states in the temporal domain. The network was initially trained within a Python based framework, after which the learned weights ($\mathrm{V_{FG}}_{0}$) were mapped onto the Verilog-A model of the FG synapse for circuit-level evaluation in Cadence.

The network output is decoded based on spike counts from the output neurons ($N_1$ and $N_2$). For each input combination, the neuron with the higher spike count determines the predicted class. Specifically, one neuron represents the logical output ‘0’ while the other represents ‘1’, enabling classification through a winner-takes-all mechanism.

The measured spike responses from Cadence simulations, shown in Fig.~\ref{fig13}, demonstrate correct XOR functionality across all input combinations. For inputs (0,0) and (1,1), the network produces higher activity in the neuron corresponding to logical ‘0’, while for (0,1) and (1,0), the neuron corresponding to logical ‘1’ dominates. This behavior is consistent with the XOR truth table, confirming correct nonlinear decision boundary formation in the mixed-signal SNN. 

Furthermore, the corresponding Python-based hardware-aware model achieves 100\% classification accuracy, with closely matching area (0.003543\,mm$^2$) and power (24.958\,$\mu$W) estimates. This agreement validates the fidelity of the proposed framework in bridging algorithm-level simulation and circuit-level implementation.

The execution variance between these two environments is notable: the Cadence simulation finished in $41.38 s$, while the Python-based testing time was approximately $268.03 ms$. These results illustrate the additional processing overhead inherent in transistor-level circuit validation.


\subsection{SNN with Neuromorphic Datasets}
Subsequently, the proposed framework is evaluated across multiple datasets and network architectures to assess its performance under both ideal and hardware-constrained conditions. Specifically, the mixed-signal SNN is tested on the \textbf{N-MNIST}, \textbf{DVS Gesture}, and \textbf{SHD} datasets using both digital (\texttt{nn.Linear}) designed using Snn Torch \cite{eshraghian_training_2023-1} and hardware-oriented synapse models (ReRAM and Floating-Gate), combined with different neuron implementations. The results are summarized in Table~\ref{tab:snn_accuracy_comparison}.

\begin{table}[htbp]
\centering
\caption{Mixed Signal Spiking Neural Network with Various Datasets}
\label{tab:snn_accuracy_comparison}
\begin{adjustbox}{width=1\textwidth}
\setlength{\tabcolsep}{5pt}
\begin{tabular} {lcccccc}
\hline
\textbf{Dataset} & \textbf{Synapse} & \textbf{Neuron} & \textbf{Architecture} & \textbf{Train Acc. (\%)} & \textbf{Test Acc. (\%)} & \textbf{Epochs} \\ 
\hline
\rowcolor{white}
N-MNIST & \texttt{nn.Linear} & \texttt{snn.Leaky} & 578 $\rightarrow$ 250 $\rightarrow$ 100 $\rightarrow$ 10 (FC) & 95.74 & 95.20 & 20 \\
\rowcolor{lightgray}
N-MNIST & \texttt{Re-RAM} & LIF(65 nm) & 578 $\rightarrow$ 250 $\rightarrow$ 100 $\rightarrow$ 10 (FC) & 88.24 &86.79 & 20 \\
\rowcolor{lightgray}
N-MNIST & \texttt{Re-RAM} & Sch-LIF (28 nm) & 578 $\rightarrow$ 250 $\rightarrow$ 100 $\rightarrow$ 10 (FC) & 88.80 &84.30 & 20 \\
\rowcolor{white}
DVS Gesture & \texttt{nn.Linear} & \texttt{snn.Leaky} & 2048 $\rightarrow$ 200 $\rightarrow$ 100 $\rightarrow$ 11 (FC) & 98.14 & 81.06 & 25 \\
\rowcolor{lightgray}
DVS Gesture & \texttt{Re-RAM} & Axon-Hillock (65 nm) & 2048 $\rightarrow$ 200 $\rightarrow$ 100 $\rightarrow$ 11 (FC) & 92.20 & 73.48 & 25 \\
\rowcolor{white}
SHD & \texttt{nn.Linear} & \texttt{snn.Leaky} & 700 $\rightarrow$ 150 $\rightarrow$ 20 (R) & 96.26 & 69.30 & 20 \\
\rowcolor{lightgray}
SHD & \texttt{Floating-Gate} & Axon-Hillock (65 nm) & 700 $\rightarrow$ 150 $\rightarrow$ 20 (R) & 90.55 & 64.66 & 35 \\
\rowcolor{lightgray}
SHD & \texttt{Floating-Gate} & Hodgkin-Huxley (65 nm) & 700 $\rightarrow$ 200 $\rightarrow$ 100$\rightarrow$ 20 (FC) & 89.12 & 66.2 & 35 \\
\rowcolor{lightgray}
SHD & \texttt{Re-RAM} & LIF(28 nm) & 700 $\rightarrow$ 200 $\rightarrow$ 100 $\rightarrow$ 20 (FC) & 91.30 & 67.50 & 35 \\
\hline
\end{tabular}
\end{adjustbox}
\end{table}

\begin{figure}[htbp]
    \centering
    \includegraphics[width=0.95\columnwidth]{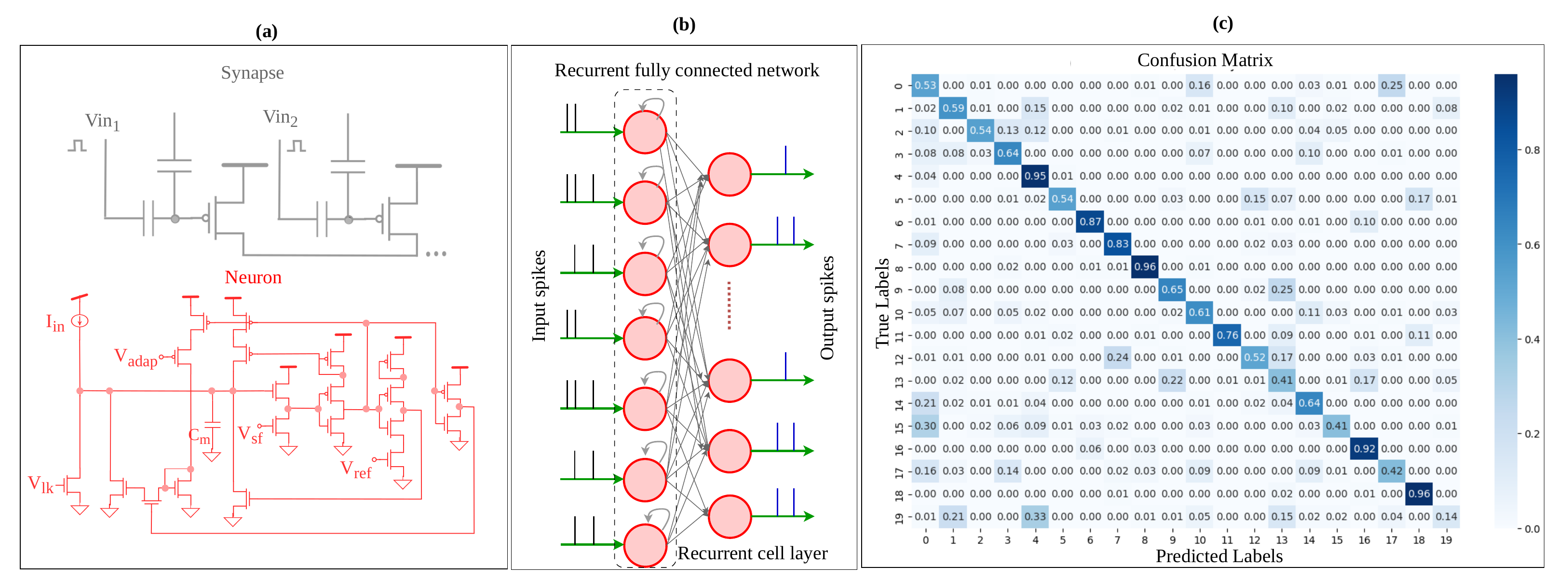}
    \caption{FG synapse-based SNN architecture for SHD dataset classification. (a) Schematic of the FG transistor-based synapse and the LIF neuron model (b) Recurrent SNN architecture with recurrent connections, where synaptic weights are implemented using FG transistors and updated via BPTT (c) Confusion matrix obtained from the classification of the SHD dataset using the SNN.}
    \label{fig14}
\end{figure}
For the \textbf{N-MNIST} dataset, the digital baseline achieves a test accuracy of \textbf{95.20\%}, while replacing ideal synapses with ReRAM-based implementations reduces the accuracy to \textbf{86.79\%} (LIF, 65\,nm) and \textbf{84.30\%} (Schmitt-LIF, 28\,nm), reflecting the impact of hardware-induced non-idealities. On the \textbf{DVS Gesture} dataset, the digital implementation achieves \textbf{81.06\%} test accuracy, whereas the ReRAM-based model using an Axon-Hillock neuron achieves \textbf{73.48\%}. For the \textbf{SHD} dataset, a temporally complex auditory benchmark, the digital recurrent baseline achieves \textbf{69.30\%} test accuracy, while hardware-aware implementations achieve \textbf{64.66\%} (FG + Axon-Hillock), \textbf{66.2\%} (FG + Hodgkin–Huxley), and \textbf{67.50\%} (ReRAM + LIF). Across all experiments, the same learning framework described earlier is used without modification; although learning with nonlinear analog synapses may benefit from specialized training algorithms, such adaptations are beyond the scope of this work. An example of an FG synapse-based SNN architecture for SHD classification is shown in Fig.~\ref{fig14}, illustrating the integration of device-level synapses with network-level learning.

In addition to classification accuracy, hardware metrics such as area and power consumption are critical for evaluating the feasibility of mixed-signal SNN deployment. Synaptic power is tracked by accumulating the product of the total output branch current and the supply voltage ($V_\mathrm{DD}$) at each timestep. The accumulated value is then normalized over the total number of timesteps to find the average power. 

\begin{equation}
P_t = (I_\mathrm{total}) \cdot V_{DD}, 
\qquad
\bar{P} = \frac{1}{N_t} \sum_{t=1}^{N_t} P_t
\end{equation}

Neuron power consumption is determined by combining continuous current drawn with discrete, event-based operational costs. The continuous component is calculated at every timestep by multiplying internal physical current sources such as leak, refractory and adaptive currents by the supply voltage ($V_\mathrm{DD}$). Added to this are fixed energy-per-event costs: spike and non-spike power values are mapped directly from the simulation characteristics depending on the neuron type. 
To produce the final metric, this total accumulated power is averaged across the batch dimension and summed across all neurons, yielding a single scalar estimate of the average power.

Similarly, the layout area of individual core blocks are mapped from Cadence simulations. Using these device-level baselines, the tool analytically computes the total area and energy footprint for the entire user-defined architecture based on component scaling and utilization. 
\begin{table}[htbp]
\centering
\small
\caption{Hardware metrics comparison of mixed-signal Spiking Neural Network
architectures using Adaptive LIF neurons (N-MNIST and SHD datasets)}
\label{tab:area_power}
\scalebox{0.85}{
\begin{tabular}{l c c c c c c }
\toprule
\textbf{Model} & \textbf{Dataset} & \textbf{Architecture} & \textbf{Synapse/Neurons} & \textbf{Area($mm^2$)} & \textbf{Power($mW)$} \\
\midrule

FC-SNN (FG) 
& N-MNIST
& $578 \times 250 \times 100 \times 10$ 
& 0.341 M
& 13.65 
& 52.07 \\

Rec-SNN (FG)
& SHD
& $700 \times 150 \times 20$  
&  0.261 M
&  10.44
& 39.59 \\

FC-SNN (ReRAM)
& N-MNIST
& $578 \times 250 \times 100 \times 10$  
& 0.184 M
& 0.107
& 20.63 \\

\bottomrule
\end{tabular}}
\end{table}

Table~\ref{tab:area_power} summarizes these metrics for representative architectures implemented with adaptive LIF neurons. The fully connected FG-based SNN achieves an accuracy of 86.79\% with an area of 13.65\,mm$^2$ and power consumption of 52.07\,mW. In comparison, the recurrent FG-based architecture reduces area and power to 10.44\,mm$^2$ and 39.59\,mW, respectively, highlighting the efficiency gains achievable through architectural optimization. The ReRAM-based implementation further reduces area (0.107\,mm$^2$) and power (20.63\,mW), demonstrating its advantage for compact and low-power neuromorphic systems, albeit with some trade-off in accuracy. These results illustrate the design trade-offs between synapse technology, architecture, and hardware efficiency.

\begin{table}[htbp]
\centering
\small
\caption{Performance of SNN architectures using Axon-hillock with different synapse and quantization levels}
\label{tab:quant}
\begin{tabular}{l c c c c c }
\toprule
\textbf{Dataset} & \textbf{Architecture} & \textbf{Synapse} & \textbf{Quantization} & \textbf{Test Accuracy} \\
\midrule

NMNIST 
& $578 \times 250 \times 100 \times 10$ (FC) 
& ReRAM 
& 3-bit 
& 81.24 \\

\midrule

SHD 
& $700 \times 150 \times 20$ (RNN) 
& \multirow{2}{*}{FG} 
& \multirow{2}{*}{8-bit} 
& 61.35 \\

DVS 
& $2048 \times 200 \times 100 \times 11$ (FC) 
& 
& 
& 70.08 \\

\bottomrule
\end{tabular}
\end{table}

The impact of synaptic quantization on SNN performance is presented in Table~\ref{tab:quant}. For the \textbf{N-MNIST} dataset, a ReRAM-based implementation with \textbf{3-bit} quantization achieves a test accuracy of \textbf{81.24\%}, reflecting the effect of limited conductance levels in practical devices. For more complex datasets, Floating-Gate (FG) synapses with \textbf{8-bit} quantization achieve test accuracies of \textbf{61.35\%} on SHD and \textbf{70.08\%} on DVS Gesture. The higher precision supported by FG devices enables improved representation of synaptic weights, particularly for temporally demanding tasks. Overall, these results highlight the sensitivity of SNN performance to synaptic precision and emphasize the importance of hardware-aware quantization in mixed-signal implementations.

\section{Discussion and Conclusion}

The proposed hardware-aware SNN framework enables quantitative evaluation of accuracy, area, power, and synaptic precision across multiple datasets and architectures. Experimental results show that replacing ideal digital synapses with hardware-realistic implementations introduces an accuracy degradation of approximately \textbf{8–11\%} on N-MNIST and \textbf{7–8\%} on DVS Gesture, primarily due to nonlinear device characteristics and limited precision, though we note that disentangling the individual contribution of each factor through a dedicated ablation study remains an important direction for future work. On the SHD dataset, which is more sensitive to temporal dynamics, the performance gap is smaller (\textbf{2–5\%}), indicating that hardware-aware models better preserve temporal information, particularly when recurrent or biologically detailed neuron models are employed.

From a hardware perspective, clear trade-offs are observed across synapse technologies. Floating-Gate (FG)-based implementations provide higher precision but incur larger area and power overheads, with the fully connected architecture occupying \textbf{13.65\,mm$^2$} and consuming \textbf{52.07\,mW}. In contrast, ReRAM-based implementations reduce area by more than \textbf{100$\times$} (down to \textbf{0.107\,mm$^2$}) and power by approximately \textbf{2.5$\times$}, making them suitable for compact and energy-efficient designs. Recurrent architectures further reduce area and power (to \textbf{10.44\,mm$^2$} and \textbf{39.59\,mW}, respectively) with minimal loss in accuracy, highlighting the impact of architectural choices under hardware constraints.

The impact of synaptic quantization is also evident. Reducing precision to \textbf{3-bit} in ReRAM-based systems results in an accuracy drop of approximately \textbf{5–7\%} on N-MNIST, whereas FG-based implementations with \textbf{8-bit} precision maintain higher accuracy across more complex datasets. This underscores the importance of synaptic precision, particularly for tasks requiring fine temporal resolution.

It is important to emphasize that this work focuses on modeling hardware-aware neuromorphic primitives rather than developing new learning algorithms. All experiments employ a unified training framework without modification to account for device nonlinearities. Despite this, stable convergence is achieved with training accuracies exceeding \textbf{88\%} across all configurations, demonstrating that the framework effectively captures device-level effects while remaining compatible with standard training approaches.

The framework is based on pulse-driven synaptic interactions and point neuron models, reflecting common mixed-signal neuromorphic implementations. This abstraction enables efficient modeling of spike-driven current accumulation while maintaining scalability, although it does not capture more complex dendritic or compartmental dynamics.

Finally, while this work does not explicitly optimize for area, power or accuracy, it provides a unified platform for evaluating these metrics consistently across architectures and technologies. The framework can be extended to support joint optimization, including network architecture search with hardware metrics incorporated into the objective function.

Overall, the results demonstrate that the proposed framework enables systematic exploration of accuracy–area–power trade-offs in mixed-signal SNNs, providing a practical pathway for translating algorithmic designs into hardware implementations with predictable performance characteristics.
\vspace{-3mm}

\section{Limitations}

The current implementation utilizes a hardware-based modeling of FG and ReRAM-based synapses, different network topologies and mixed-signal neurons. By extracting physical parameters into a Python-based environment, the tool provides a predictive analysis of classification accuracy, area and power consumption (pertaining to different datasets). 

While our Python model successfully approximates hardware-like behavior, there are inherent challenges in mapping discrete mathematical abstractions to physical analog dynamics. Specifically, the continuous evolution of the membrane potential and the spike timings in hardware may exhibit subtle non-linearities (as depicted in the supplementary information) that are difficult to capture fully within a simulated environment .

Additionally, the peripheral system-level modelling remains reserved for future implementation. Considering a real-time neuromorphic chip modeling, requires the Address Event Representation (AER) (as modeled in SANA-FE\cite{boyle2025sana}) to enable efficient, event-driven communication across layers. While the current tool explicitly models the local neuron dynamics and capture the true inter-spike interval; transitioning to a physical chip deployment (and modeling) introduces routing as well as timing constraints. The AER relies on precise event arbitration and bounded latency, any simulated delay between neuron-to-neuron communication must be strictly managed to preserve the temporal fidelity of spike propagation. Accounting for these hardware constraints specifically inter-spike interval accuracy, arbitration delays and handshaking overhead falls outside the scope of the current tool and will be systematically addressed in future work.



Additionally, the framework needs expanding beyond idealized functional blocks; detailed analysis of analog non-idealities such as device noise, process variations(\cite{11412426}), mismatch effects, parasitics and temporal drift. Finally, the current framework does not include closed-loop calibration, adaptive compensation or circuit-level feedback mechanisms that may be required in practical hardware deployment. Such techniques are often necessary to compensate for weight-programming errors and array-level mismatch. Future work will extend this environment by integrating circuit-level calibration, noise-aware modeling, variability analysis, and tighter co-simulation between Python-based models and network-level hardware realizations to maximize hardware fidelity.
\section{Code}
The preliminary tool is available at the following gitlab repo. \href{https://code.umd.edu/sshah389/mixed-signal-snns}{Mixed-signal-SNNs}

\vspace{-3mm}
\section{Acknowledgements}
The Study is Supported by Lockheed Martin University Engagement Funds. Sayma Nowshin Chowdhury is funded via DEVCOM Army Research Laboratory under Cooperative Agreement Number W911NF-24-2-0048 starting January 2024.

\section*{References}
\bibliographystyle{IEEEtran}
\bibliography{./references1.bib,
./references_2.bib}


\renewcommand{\thefigure}{S\arabic{figure}}
\renewcommand{\thetable}{S\arabic{table}}
\renewcommand{\theequation}{S\arabic{equation}}
\setcounter{figure}{0}
\setcounter{table}{0}

\title{Supplementary Material }
\maketitle

\section{Neuron dynamics}
\subsection{Standard LIF neuron}

\begin{figure}[ht]
\begin{center}
 \epsfig{file=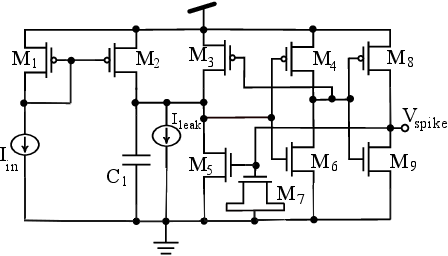, width=0.65\columnwidth}
\end{center}
\caption{Standard LIF neuron circuit used in the manuscript}
\label{figS1}
\end{figure}

The LIF neuron circuit, shown in Fig. \ref{figS1}, uses a membrane capacitor ($C_1$) to integrate input current ($I_\mathrm{in}$). Its temporal dynamics are governed by:
\begin{equation}
C_1 \frac{dV_\mathrm{mem}}{dt} = -g_\mathrm{l} \big(V_\mathrm{mem} - V_\mathrm{reset}\big) + I_\mathrm{in}(t)
\end{equation}

where $g_\mathrm{l}$ represents the leakage conductance and $V_\mathrm{reset}$ is the reset potential. As $I_\mathrm{in}$ charges $C_1$, the membrane potential ($V_\mathrm{mem}$) rises until it hits a threshold, triggering a spike via the inverter ($M_8$–$M_9$). This spike activates reset transistor $M_5$, discharging the capacitor to $V_\mathrm{reset}$ to conclude the firing event.



\newpage

\subsection{Schmitt LIF neuron}

\begin{figure}[ht]
\begin{center}
 \epsfig{file=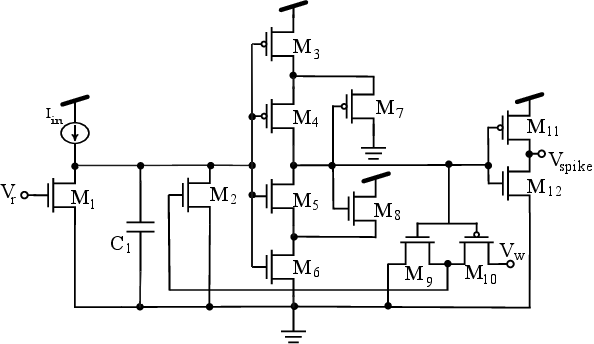, width=0.8\columnwidth}
\end{center}
\caption{Schmitt LIF neuron circuit used in the manuscript}
\label{figS2}
\end{figure}
The Schmitt Trigger-based LIF neuron [Fig. \ref{figS2}] integrates input current $I_\mathrm{in}$ onto capacitor $C_\mathrm{1}$. The membrane potential $V_\mathrm{mem}$ evolves as:$$V_\mathrm{mem}(t) = I_\mathrm{in}(t)R + (V_\mathrm{out} - I_\mathrm{in}(t)R)e^{-t/\tau}$$When $V_\mathrm{mem}$ reaches the threshold, the Schmitt Trigger activates, triggering an inverted feedback signal to transistor $M_2$. This rapidly discharges the capacitor to $V_\mathrm{reset}$ and generates an output spike, with the pulse width modulated by $V_\mathrm{w}$. Leakage is controlled by transistor $M_1$ via $V_\mathrm{r}$, providing a continuous discharge path. Repetitive spiking occurs as long as $I_\mathrm{in}$ exceeds the leakage current. 

\subsection{Temporal Dynamics Analysis}
\begin{figure*}[!htbp]
    \centering

    \begin{subfigure}[t]{0.8\textwidth}
        \centering
        
  \epsfig{file=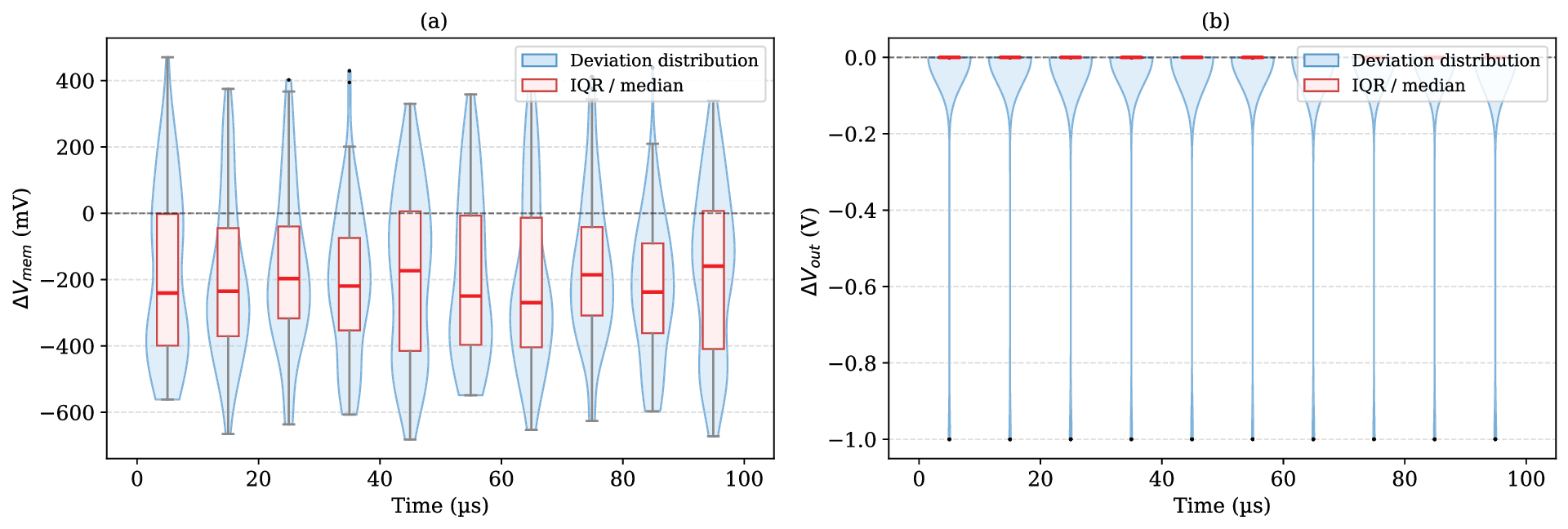, width=\columnwidth}
    \end{subfigure}

    \begin{subfigure}[t]{0.8\textwidth}
        \centering
         \epsfig{file=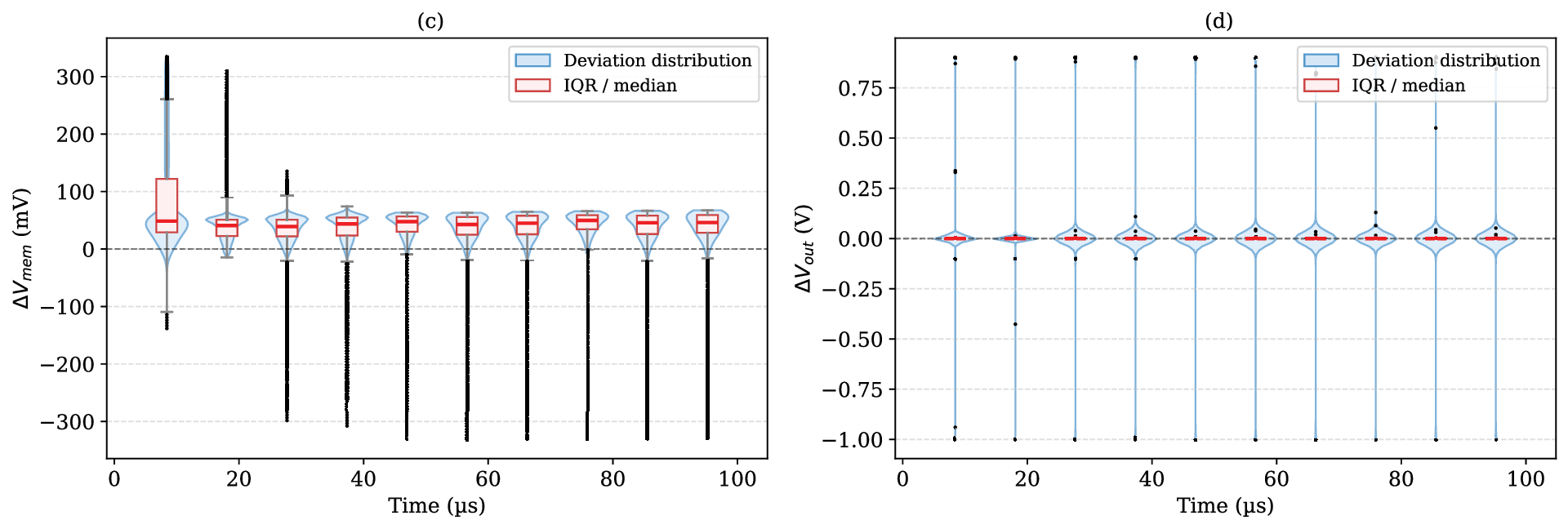, width=\columnwidth}
    \end{subfigure}

    \begin{subfigure}[t]{0.8\textwidth}
        \centering
         \epsfig{file=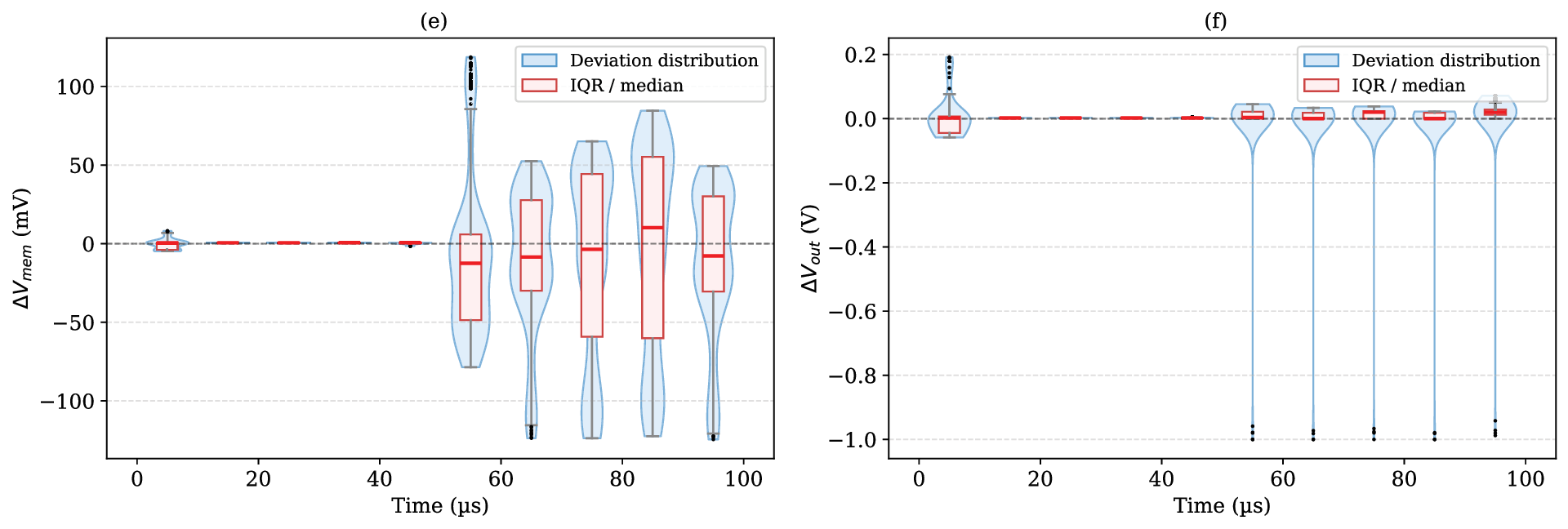, width=\columnwidth}
    \end{subfigure}

    \begin{subfigure}[t]{0.8\textwidth}
        \centering
         \epsfig{file=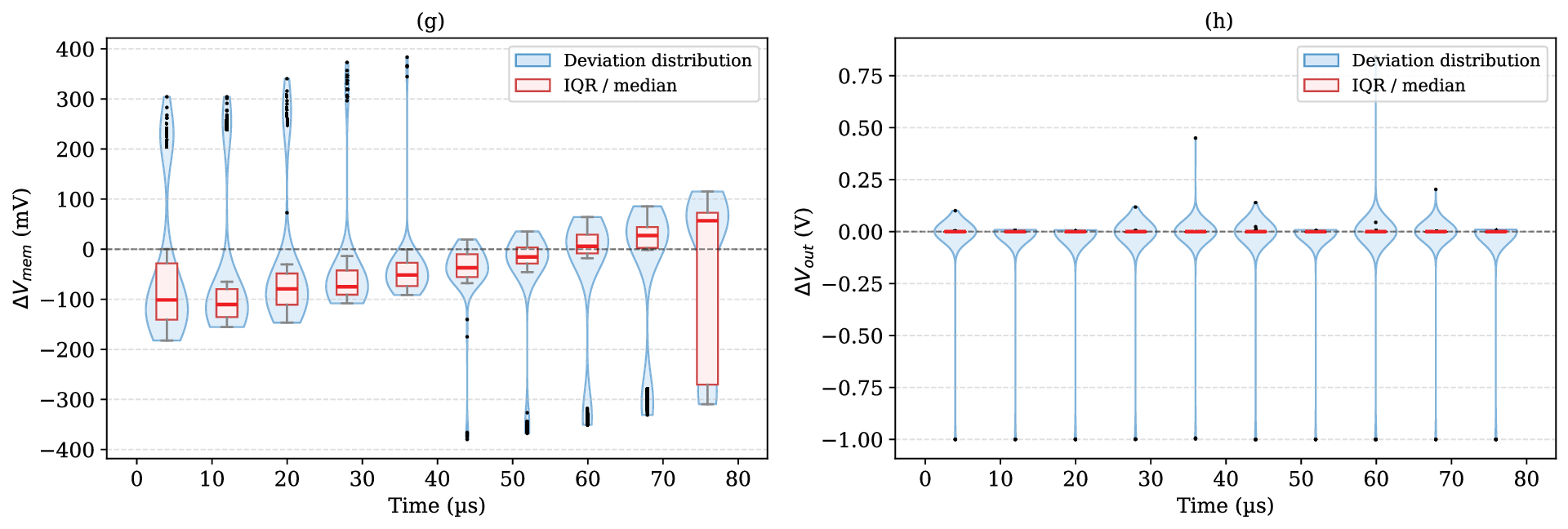, width=\columnwidth}
    \end{subfigure}

    \begin{subfigure}[t]{0.48\textwidth}
        \centering
        \epsfig{file=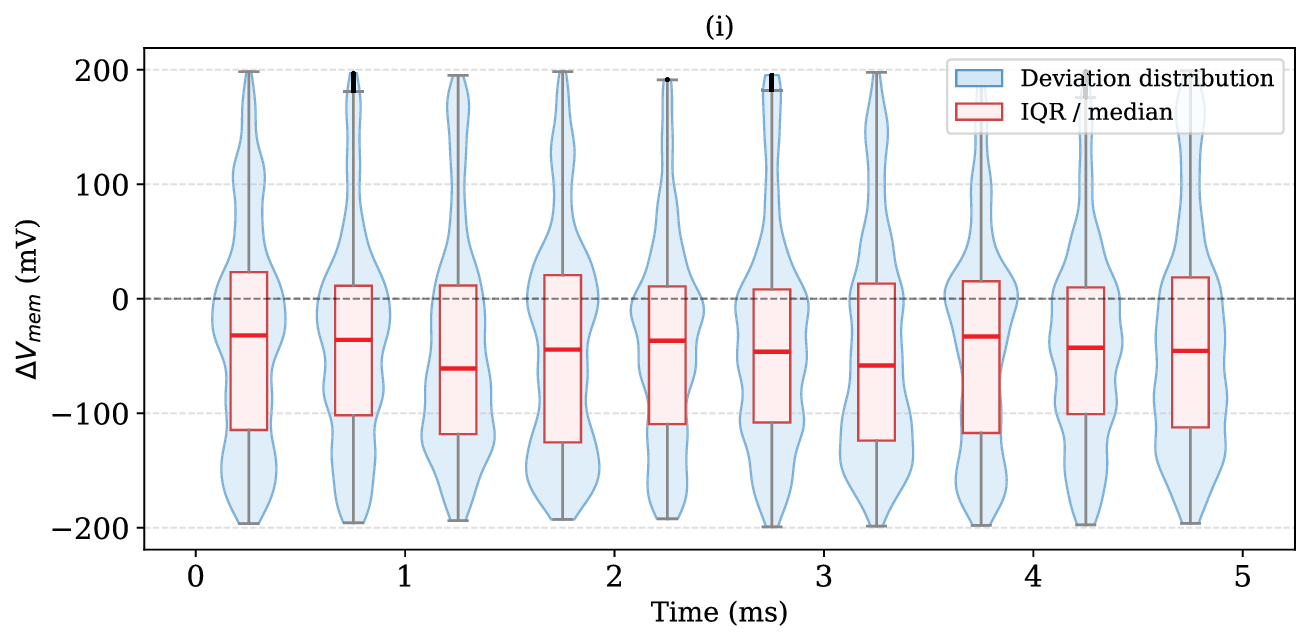, width=\columnwidth}
    \end{subfigure}
    \caption{Temporal deviation between Cadence circuit-level simulation and Python models for neuron architectures: (a)-(b): Standard LIF ,(c)-(d): Schmitt-trigger LIF, (e)-(f): Axon Hillock, (g)-(h) Adaptive LIF membrane potential and spike output deviation and (i) Hodgkin--Huxley model membrane potential deviation}
   
    \label{figS3}
\end{figure*}

To validate the analytical Python models against circuit-level Cadence simulations, the deviation was computed across the full simulation duration for each neuron architecture. The deviation was evaluated at every timestep and aggregated into ten equal time windows, with the resulting distribution characterised by the interquartile range (IQR) and median across each window (as represented in Fig. \ref{figS3}). 

This analysis was performed under a fixed synaptic current ($I_{syn}$) for each architecture, examining both the membrane potential $V_{mem}$ and the spike output $V_{out}$ for the Axon Hillock neuron ($I_{syn}$ = 20 nA), Adaptive LIF neuron ($I_{syn}$ = 50 nA), Standard LIF neuron ($I_{syn}$ = 100 nA), Schmitt-trigger LIF neuron ($I_{syn}$ = 500nA) and for $V_{mem}$ in the case of the Hodgkin--Huxley neuron ($I_{syn} = 20 nA$). 

The $\Delta V_{mem}$  distributions show moderate deviations concentrated at spike edges, while the inter-spike intervals exhibit closer agreement (resulting minimum deviation in $\Delta V_{out}$), reflecting that the Python models capture the overall spiking rhythm and steady-state firing rate across architectures. 

Across all five neuron models, the median deviation in $\Delta V_{\mathrm{out}}$ remains close to zero throughout the simulation duration, while $\Delta V_{\mathrm{mem}}$ displays moderate distributional spread, particularly during active spiking regions. The spread of the distribution captures the variability introduced by spike transients; brief moments during threshold crossing where the two signals diverge due to sub-timestep timing differences and transistor-level non-linearity. These deviations remain consistent across temporal windows, suggesting that the mismatch is governed by circuit-level non-idealities rather than any cumulative drift in the analytical model.

\section{Quantization}
\subsection{Floating gate synapse}

\begin{figure}[ht]
\begin{center}
\epsfig{file=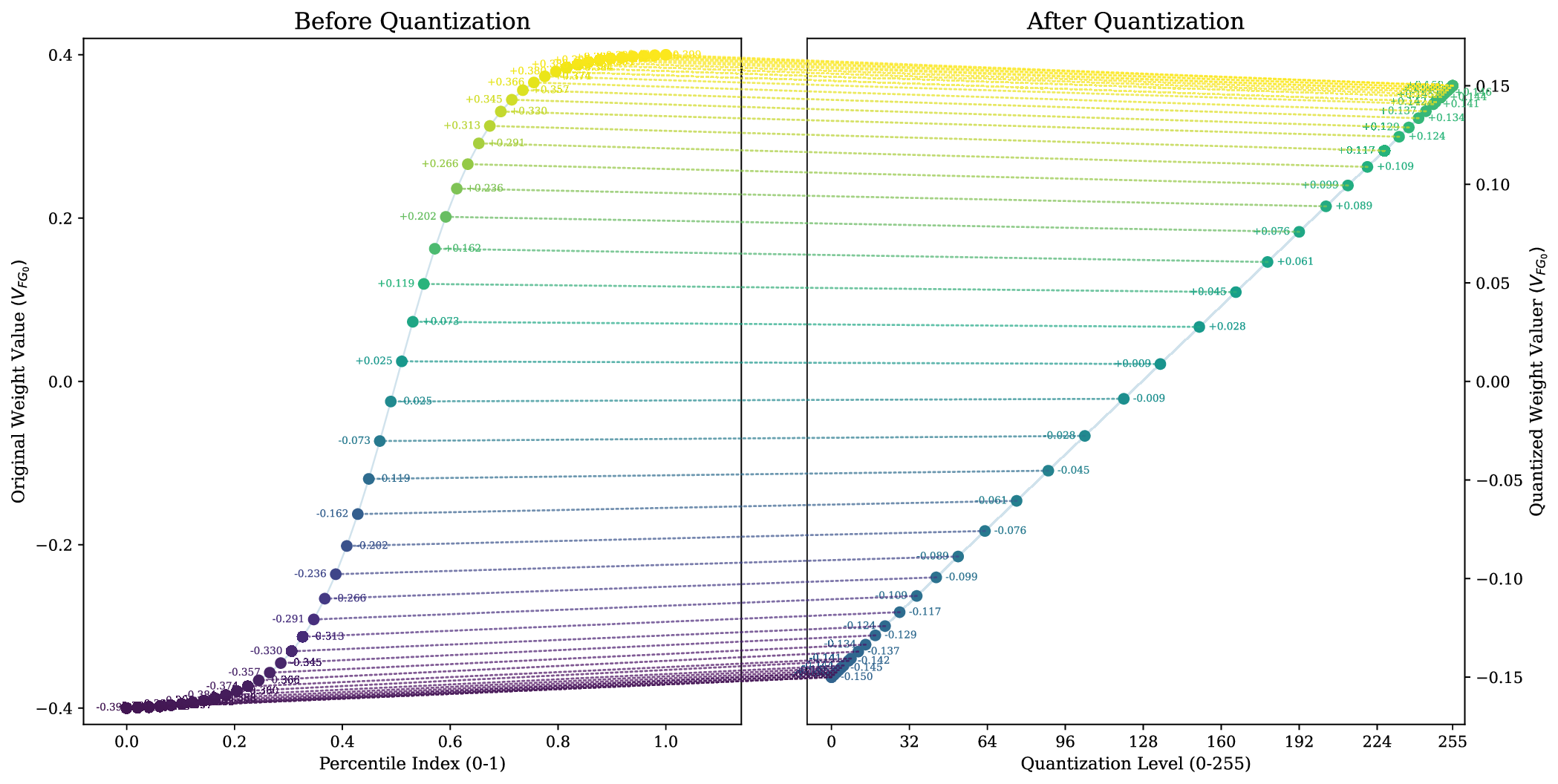, width=0.9\columnwidth}
\end{center}
\caption{Discrete Weight Mapping via 8-bit Post-Training Quantization of the Floating-Gate Synapse}
\label{figS8}
\end{figure}

This work implements a hardware-grounded 8-bit post-training quantization (PTQ) scheme rooted in the floating-gate synapse characterization frameworks established by \cite{chowdhury2024analysis,chen_open-source_2024}. As characterized, the physical device exhibits threshold voltage ($V_{\text{tp}}$) levels distributed across a $0\text{–}5\text{ V}$ dynamic range with device-level voltage resolution of approximately $19.61\text{ mV}$ per step. The corresponding operating drain current ($I_d$) spans a range of $10^{-8}\text{ A}$ to $10^{-4}\text{ A}$. Informed by these hardware constraints, the network's continuous weight parameters ($V_{FG_0}$ $\in$ [-0.4,\,+0.4]) are mapped directly onto $2^8 = 256$ discrete quantization levels , yielding an effective weight representation range of $V_{FG_0}$ $\in$ [-0.15,\,+0.15]. The resulting weight quantization mapping behavior is illustrated in the Fig. \ref{figS8}.

\subsection{ReRAM synapse}
\begin{figure}[ht]
\begin{center}
\epsfig{file=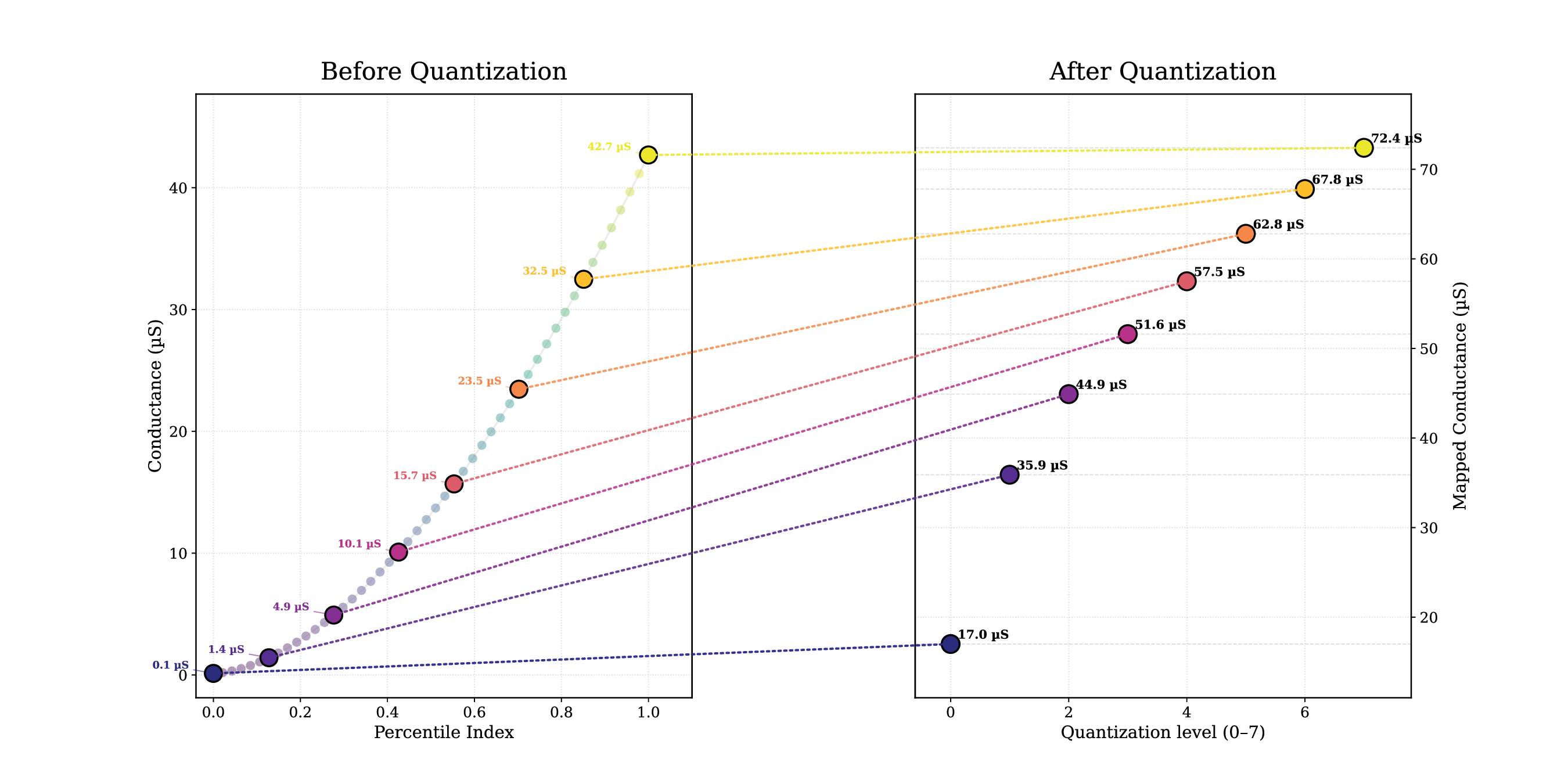, width=0.9\columnwidth}
\end{center}
\caption{Discrete Weight Mapping via 3-bit Post-Training Quantization of the ReRAM Synapse}
\label{figS9}
\end{figure}

Additionally, this work also implements a 3-bit PTQ scheme for the ReRAM synapses grounded in the device characterisation framework established in \cite{didin2026characterization}. Pertaining to these exact hardware parameters (measured conductances $\in$ [17.0\ $\mu$\text{S},\,72.4\ $\mu$\text{S}]), the network's continuous weight parameters ($G$) are mapped directly onto $2^3 = 8$ discrete conductance levels. The resulting ReRAM weight quantization mapping behavior is shown in Fig. \ref{figS9}.

\end{document}